\let\accentvec\vec
\documentclass{aa}
\let\vec\accentvec
\usepackage{graphicx}
\usepackage{graphicx}

\usepackage{natbib}
\usepackage{amssymb}
\usepackage{amsmath}
\usepackage{xspace}
\usepackage{color}
\usepackage{mathrsfs}
\usepackage{xcolor}

\usepackage[varg]{txfonts}
\usepackage[flushleft]{threeparttable}
\newcommand{\be}{\begin{equation}}
\newcommand{\ee}{\end{equation}}

\begin{document} 

   \title{Stochastic re-acceleration and magnetic-field damping in Tycho's supernova remnant}

   \author{A. Wilhelm\inst{1,2}\fnmsep\thanks{Corresponding author, \email{alina.wilhelm@desy.de}}
          \and
	  I. Telezhinsky\inst{1,2}
          \and
	  V.V. Dwarkadas\inst{3}
          \and
          M. Pohl\inst{1,2}}

\institute{DESY, 15738 Zeuthen, Germany 
\and Institute of Physics and Astronomy, University of Potsdam, 14476 Potsdam, Germany
\and University of Chicago, Department of Astronomy\& Astrophysics, 5640 S Ellis Ave, ERC 569, Chicago, IL 60637, U.S.A.}

\date{Received ; accepted }

 
  \abstract
   {Tycho's supernova remnant (SNR) is associated with the historical supernova (SN) event SN~1572 of Type Ia. The explosion occurred in a relatively clean environment, and was visually observed, providing an age estimate. Therefore it represents an ideal astrophysical test-bed for the study of cosmic-ray acceleration and related phenomena. A number of studies suggest that shock acceleration with particle feedback and very efficient magnetic-field amplification combined with Alfv\'{e}nic drift are needed to explain the rather soft radio spectrum and the narrow rims observed in X-rays.}
   {We show that the broadband spectrum of Tycho's SNR can alternatively be well explained when accounting for stochastic acceleration as a secondary process. The re-acceleration of particles in the turbulent region immediately downstream of the shock should be efficient enough to impact particle spectra over several decades in energy. The so-called Alfv\'{e}nic drift and particle feedback on the shock structure are not required in this scenario. Additionally, we investigate whether synchrotron losses or magnetic-field damping play a more profound role in the formation of the non-thermal filaments.}
   {We solve the full particle transport equation in test-particle mode using hydrodynamic simulations of the SNR plasma flow. The background magnetic field is either computed from the induction equation or follows analytic profiles, depending on the model considered. Fast-mode waves in the downstream region provide the diffusion of particles in momentum space.}
   {We show that the broadband spectrum of Tycho can be well explained if magnetic-field damping and stochastic re-acceleration of particles are taken into account. Although not as efficient as standard diffusive shock acceleration (DSA), stochastic acceleration leaves its imprint on the particle spectra, which is especially notable in the emission at radio wavelengths. We find a lower limit for the post-shock magnetic-field strength $\sim330\,\mathrm{\mu G}$, implying efficient amplification even for the magnetic-field damping scenario. For the formation of the filaments in the radio range magnetic-field damping is necessary, while the X-ray filaments are shaped by both the synchrotron losses and magnetic-field damping.}
   {}

   \keywords{Acceleration of particles -- cosmic rays : supernova remnants -- ISM}

   \maketitle
%

\section{Introduction}

Supernova remnants (SNRs) are among the most exciting astrophysical objects, because they may shed light on a major question of physics: the
origin of cosmic rays. An interesting object in this respect is the remnant of the historical Type Ia event SN~1572, first described by Tycho Brahe.
Due to his observation, the age of Tycho's SNR (SNR G120.1+1.4, hereafter Tycho) is accurately determined to be $\sim$ 440 years. Tycho originated from a Type Ia supernova (SN) \citep{2008Natur.456..617K}, and is assumed to have a canonical explosion energy of $\sim10^{51}$~erg. These details, together with the available broadband spectrum, make Tycho 
one of the best astrophysical laboratories to study particle acceleration. Nevertheless, 
several questions still remain unsolved.

Indeed, despite the success of diffusive shock acceleration (DSA) theory, it fails to
explain the observed radio spectrum $S_\nu$. The measured
radio spectral index $\sigma \approx$ 0.65, with $S_\nu \propto \nu ^{-\sigma}$ \citep{2006A&A...457.1081K}, significantly deviates from the standard DSA prediction $\sigma \approx$ 0.5.
This discrepancy is generally accounted for using the concept of Alfv\'{e}nic drift \citep{1978MNRAS.182..147B}.
Hence, various authors postulate Alfv\'{e}nic drift only in the upstream region \citep{2008A&A...483..529V, 2012A&A...538A..81M}, or in the upstream
and downstream \citep{2014ApJ...783...33S} regions of Tycho's forward shock. The proper motion of cosmic-ray scattering centers that proceed with Alfv\'en speed is assumed to decrease the compression ratio felt by the particles and therefore
to cause a softening of their spectra. To be more precise, the effective compression ratio seen by particles is
\begin{equation}
 r_{\mathrm{eff}}=\frac{u_1+H_{c1}v_{\mathrm{A}1}}{u_2+H_{c2}v_{\mathrm{A}2}}\,,
 \label{eq:comp_ratio}
\end{equation}
where $u_1$, $v_{\mathrm{A}1}$ are the plasma velocity and the Alfv\'en velocity in the upstream and $u_2$,
$v_{\mathrm{A}2}$ in the downstream regions, respectively. The relative direction of the propagation of the Alfv\'en waves in the upstream (downstream) is
reflected by the cross helicity, $H_\mathrm{c1(2)}$. In the above global models its value is chosen rather freely\footnote{Usually it is taken $H_\mathrm{c1}=-1$ and
$H_\mathrm{c2}=0$ \citep{2008A&A...483..529V, 2012A&A...538A..81M} or $H_\mathrm{c1}=-1$ and $H_\mathrm{c2}=1$ \citep{2014ApJ...783...33S}.},
in order to reduce the effective compression ratio and thereby to account for the soft particle spectrum. On closer inspection, however, the concept of Alfv\'{e}nic drift as an explanation for the spectral softening emerges as misleading.

First of all, \cite{1999A&A...343..303V} performed a detailed calculation on self-consistent transmission of the Alfv\'en waves through
the shock and found that the presence of waves results in a harder particle spectra than predicted by the standard
theory (meaning that Alfv\'en waves infer exactly the opposite effect as claimed in \cite{2008A&A...483..529V}, \cite{2012A&A...538A..81M} and \cite{2014ApJ...783...33S}). The reason is that Alfv\'en waves that move in the upstream region in the opposite direction as the background plasma
($H_\mathrm{c1}=-1$), propagate also predominately in the opposite direction in the downstream ($H_\mathrm{c2}\approx -1$)\citep{1999A&A...343..303V}. Therefore, even despite the modifications induced by the strong
magnetic-field pressure, the effective compression ratio seen by particles (Eq.~\ref{eq:comp_ratio}) exceeds the standard strong-shock value ($r_\mathrm{eff}>4$).
Investigating the impact of the Alfv\'enic drift within the framework of \cite{1999A&A...343..303V} we find that for the Alfv\'enic Mach numbers, $M_\mathrm{A}\equiv u_1/v_{1,\mathrm{A}}$, in the range 10-13 (as presented in \cite{2008A&A...483..529V} and \cite{2012A&A...538A..81M}) the particle spectral index for a strong shock results in $s\approx 1.9$ instead of the $s\approx2.3$, required by radio observations of Tycho. Furthermore, the negative downstream helicity, $H_\mathrm{c2}\approx -1$, predicted by \cite{1999A&A...343..303V} is exactly reversed scenario as assumed by \cite{2014ApJ...783...33S}, who used $H_\mathrm{c2}=1$.   

Secondly, the Alfv\'enic-drift phenomenon in the global models of Tycho is often referred to result from the non-resonant streaming instability of cosmic rays \citep{BellsInst}. According to \cite{BellsInst}, the phase speed of the non-resonant modes is negligible compared to the shock velocity. But in case of Alfv\'{e}nic drift, the Alfv\'{e}n velocity required to account for the Tycho's radio spectra has to be enormous.

The following estimation should demonstrate the corresponding discrepancy. Let us assume that Alfv\'{e}nic drift occurs only in the upstream region ($H_\mathrm{c2}=0$ and $H_\mathrm{c1}=-1$), as in  \cite{2008A&A...483..529V} and \cite{2012A&A...538A..81M}, although it contradicts the findings of \cite{1999A&A...343..303V}. In this case, we can rearrange Eq.~\ref{eq:comp_ratio} to
\be
M_{\mathrm{A}}=(1-r_\mathrm{eff}/r_\mathrm{sh})^{-1}\,,
\label{eq:alv_vel}
\ee
where $M_\mathrm{A}\equiv u_1/v_{1,\mathrm{A}}$ is the Alfv\'{e}nic Mach number and $r_\mathrm{sh}\equiv u_1/u_2$ is the gas compression ratio of the shock.
The sub-shock compression ratio in \cite{2008A&A...483..529V} and \cite{2012A&A...538A..81M} is in the range $r_\mathrm{sh}=3.7-3.9$, even if the non-linear effects of the DSA are included. The effective compression ratio, seen by particles, required by the radio observations is $r_\mathrm{eff}\approx 3.3$. Inserting these values into Eq.~\ref{eq:alv_vel} provides a relatively low Alfv\'{e}nic Mach number, $M_\mathrm{A}=6-9$, making clear that the Alfv\'{e}n speed exhibits a significant fraction of the shock velocity. Thus, to explain Tycho's radio data with Alfv\'{e}nic drift, the Alfv\'{e}n phase speed has to attain  $11\%-16\%$ of the shock velocity. Obviously this value is in conflict with the phase speed of the non-resonant mode, $v_\phi \approx 0$, as described by \cite{BellsInst}. Therefore, Alfv\'{e}nic drift in the models for Tycho
cannot be associated with the non-resonant streaming instability.

It is important to note here that \cite{2008A&A...483..529V} and \cite{2012A&A...538A..81M} applied in their modeling somewhat smaller magnetic-field values than required to best match the radio data with Alfv\'{e}nic drift. The post-shock magnetic fields of $\sim 300 \, \mathrm{\mu G}$ \citep{2012A&A...538A..81M} and $\sim 400 \, \mathrm{\mu G}$ \citep{2008A&A...483..529V} provide sufficient flux in the radio range, but fit the spectral shape of the observed data only moderately well. The Alfv\'{e}n velocity in these models (with corresponding Alfv\'{e}nic Mach numbers of $M_\mathrm{A}\approx 13$ and $M_\mathrm{A}\approx 10$) is still $\sim10\%$ of the shock velocity. The  phase speed of the non-resonant mode is much less than that and can not support Alfv\'enic drift, of course.

A post-shock magnetic field above $\sim300\,\mathrm{\mu G}$, when combined with the relatively low ambient density of $0.3-0.4\, \mathrm{cm^{-3}}$  \citep{2008A&A...483..529V,2012A&A...538A..81M}, becomes dynamically important. The corresponding magnetic-field pressure will affect the shock compression ratio, which results in $r_\mathrm{sh}<3.9$. Aside from the work of \cite{2012A&A...538A..81M}, this effect has been neglected in global models for Tycho.

The above arguments illustrate that the Alfv\'enic-drift concept is problematic to explain Tycho's soft radio spectrum consistently. An alternative way to explain the softening of the particle spectra in collisionless shocks is accounting for neutral hydrogen in the surrounding medium, first proposed by \cite{2012ApJ...755..121B}. Analytic calculations from \cite{2012ApJ...758...97O}
plus later simulations \citep{2016ApJ...817..137O} show that neutrals can leak from the downstream into the upstream region and modify the shock structure. This results in a steeper particle spectrum
as produced by standard DSA. \cite{MorBla_NH} build on that idea to model the rather soft
$\gamma$-ray spectrum of Tycho. However, the leakage of neutral particles is significant for shock velocities $V_\mathrm{sh}<3000\,\mathrm{km\,s}^{-1}$ \citep{2012ApJ...758...97O}, which is considerably below the value ascertained for Tycho. \cite{MorBla_NH} argued that certain regions of Tycho can feature slower shocks that propagate into dense, partially neutral material. The regions with slower shock velocities would have to provide nearly all of the observed overall emission, while contributions from the regions with the fast shock velocities would have to be weak, otherwise the integrated emission would reflect the hard spectra expected for fast shocks in an ionized medium.

In this work we suggest a new approach: besides standard DSA, we consider an additional acceleration process, namely the stochastic re-acceleration of particles in the immediate post-shock region of the SNR. It has been shown that fast-mode waves that survive the transit-time damping by the background plasma are efficient modes to accelerate charged particles via cyclotron resonance~\citep{Yan&Laz2002, 2008ApJ...683L.163L}. \cite{2015A&A...574A..43P} demonstrated that particles may be stochastically re-accelerated by fast-mode turbulence, which occurs in the downstream region, after escaping from the forward shock.
In this work we build on that idea and include it into a detailed modeling of Tycho.

Fast-mode turbulence may arise behind the shock from velocity fluctuations of the plasma flow, via e.g shock rippling \citep{2007ApJ...663L..41G}, and build a thin turbulent region behind the blast wave. The width of this zone is regulated by the energy transfer from the background plasma into turbulence as well as damping induced by the re-acceleration of cosmic rays. 
We show that fast-mode turbulence that carries a few percent of the energy density of the background plasma in the downstream region is strong enough to modify the spectrum of particles that have already been accelerated by the shock.
Thus, in our treatment stochastic acceleration and DSA operate together and produce a particle spectrum consistent with the observed radio spectral index. An additional advantage of our approach is that it is fully time-dependent.\footnote{Time-dependent hydrodynamics for Tycho was already used by \cite{2014ApJ...783...33S}. In their approach, however, a steady-state solution for DSA was used to numerically inject a specific cosmic-ray spectrum at the location of the forward shock.}
We solve the \textit{time-dependent transport equation} for cosmic rays that contains a DSA term and diffusion in momentum space, and is coupled to hydrodynamic simulations.   

Another interesting question regarding Tycho is to what extent the magnetic field is amplified inside the remnant. A relatively high post-shock magnetic field, $300 -400 \, \mathrm{\mu G}$, is postulated by several global models \citep{2008A&A...483..529V,2012A&A...538A..81M}.
One of the reasons is the afore-mentioned Alfv\'{e}nic drift, which demands rather large magnetic-field values to account for the radio spectrum. Since we do not postulate any Alfv\'{e}nic drift in our model, our approach of inferring the magnetic-field strength is an alternative to that of previous works on Tycho.

A major argument for a high magnetic field in the downstream of Tycho are the observed narrow non-thermal X-ray filaments \citep{2002ApJ...581.1101H, 2006A&A...453..387P, 2012SSRv..173..369H}.
Since electrons can only propagate for a finite distance before they lose their energy due to synchrotron radiation, the rim widths may reflect
the magnetic-field strength in the immediate downstream of the SNR.
An alternative scenario is provided by damping of the turbulent magnetic field in the interior of the remnant
\citep{2005ApJ...626L.101P,2014ApJ...790...85R,2015ApJ...812..101T}, in which the narrowness of the
non-thermal rims can be explained by the damping of the turbulently amplified magnetic field. For Tycho, a distinction between these scenarios by means of the energy-dependence of X-ray filaments is difficult \citep{2012A&A...545A..47R, 2015ApJ...812..101T}. Magnetic-field damping is widely considered as the scenario that allows for a weak
magnetic-field strength inside SNRs.
Nevertheless, \cite{2015ApJ...812..101T} find that in either case the minimum downstream magnetic-field value inferred from the Tycho's non-thermal
filaments is at least $\sim 20\, \mathrm{\mu G}$. Assuming that the electron acceleration is limited by the age of the remnant, 
the work from $NuStar$ collaboration \citep{2015ApJ...814..132L} estimates $\sim 30\, \mathrm{\mu G}$ for the downstream magnetic field. However, the majority of studies
cited above favor the loss-limited interpretation for particle acceleration in Tycho. Furthermore, the most realistic limit is obtained from an analysis of the entire spectral energy distribution (SED),
which provides the minimal post-shock magnetic field value of  $\sim 80 \, \mathrm{\mu G}$ \citep{2011ApJ...730L..20A}. Attempts to simultaneously fit the radio and the $\gamma$-ray data infer that, any weaker magnetic field would cause an overproduction
of $\gamma$-ray photons generated via inverse-Compton scattering. Therefore, the question about the magnetic field value is automatically tied to the question of whether Tycho's $\gamma$-ray
emission has a predominately leptonic or hadronic origin.
The hadronic scenario has been strongly favored in the literature \citep{2012A&A...538A..81M, 2013MNRAS.429L..25Z, 2013ApJ...763...14B, 2014NuPhS.256...89C,2014ApJ...783...33S},
as opposed to a leptonic model \citep{2012ApJ...749L..26A}.

For our modeling, we start from the minimal magnetic field compatible with the entire SED. The evolution of the SNR, which is computed using hydrodynamical 
simulations, occurs in a medium with a constant density.
We explicitly model the acceleration of each particle species in the test-particle limit, taking shock acceleration as well as stochastic
acceleration in the downstream region into account, with both acceleration processes being time-dependent. We explicitly model advection and diffusion of cosmic rays and take 
synchrotron losses for electrons into account. Furthermore, we consider the non-thermal radio and X-ray filaments and investigate whether they arise from extensive synchrotron losses or magnetic-field damping. For the study of the X-ray filaments it is especially important to include diffusion of the particles, otherwise the distance that they propagate and thus the rim width would be underestimated. 

This paper is organized as follows:
Section~\ref{sec:modeling} describes our method: hydrodynamical picture, magnetic profiles as well as particle acceleration via DSA and Fermi II
processes. In Section~\ref{moda} we examine the model with the minimal amplified magnetic field compatible with the $\gamma$-ray observations.
We justify the necessity for magnetic field damping that we introduce in Section~\ref{subsec:med_mag_fl}.
Therein we discuss our favorite model for Tycho, and deduce a new theoretical minimum for the magnetic-field strength based on investigation
of the multi-frequency spectrum along with non-thermal filaments. The $\gamma$-ray spectrum of the resulting model comprises both hadronic and leptonic components.
In Section~\ref{modc} we discuss the potential for a purely hadronic scenario with a strongly amplified magnetic field. 
\section{Modeling}
\label{sec:modeling}
\subsection{Hydrodynamics}
\label{sec:hyd_mf}

To accurately calculate the acceleration of particles to relativistic
energies and the high-energy emission from the remnant, we start with
an evolutionary model of the remnant. This model provides the
dynamical and kinematic properties of the remnant as a function of
time. We model the hydrodynamic evolution of the SN shock wave in
Tycho using the VH-1 code, a 1, 2 and 3 dimensional
finite-difference code that solves the hydrodynamic equations using
the Piecewise Parabolic Method of \citet{cw84}. \citet{dc98} showed
that the ejecta structure of Type-Ia SNRs such as Tycho could be best
approximated by an exponential density profile. Using this profile,
the parameters needed to model the SNR are the explosion energy of the SN, the ejected mass, and the density of the medium into which the supernova is expanding. We use the canonical value of 10$^{51}$ ergs for the energy of the explosion.  Type Ia's are presumed to arise from the thermonuclear deflagration and detonation of a white dwarf, so we assume an ejecta mass of 1.4 $M_{\odot}$, appropriate for a Chandrashekhar-mass white dwarf. The remaining parameter necessary to model the evolution is the density of the ambient medium.

Although the density around Tycho varies~\citep{2013ApJ...770..129W}, as expected for such an extended structure, the remnant appears to expand in a clean environment without any large inhomogeneities, such as molecular clouds~\citep{Tian_2011}. Therefore, an explosion in a medium with a constant density provides a reasonable description for the dynamics of the remnant. The average ambient density varies in the literature, depending on how it was measured. X-ray measurements of the expansion rate suggest an ambient density of 0.2-0.6~$\mathrm{cm^{-3}}$~\citep{Hughes_2000}. Later observations infer an upper limit of 0.2~$\mathrm{cm^{-3}}$~\citep{Katsuda_2010}. X-ray observations of~\cite{bw_Cas_Chen} reveal a lack of thermal emission in the post-shock region of Tycho, inferring an ambient density below 0.6~$\mathrm{cm^{-3}}$. A low density of 0.2~$\mathrm{cm^{-3}}$ around Tycho is obtained by~\cite{2013ApJ...770..129W}, who determined the post-shock temperature from mid-infrared emission of the remnant. On another hand, efficient particle acceleration in SNRs can reduce the downstream temperature of the plasma~\citep{Drury2009}, leading to a suppression of thermal emission and hence an underestimation of the ambient density. Higher values for the ambient density are additionally supported by \cite{dc98}, who found that densities in the range of 0.6-1.1~$\mathrm{cm^{-3}}$ better matched the observations of Tycho. Furthermore, \cite{Kozlova_2018} and \cite{Badenes_2006} favor a delayed detonation model with $\sim1.0\,\mathrm{cm^{-3}}$ for the X-ray morphology of Tycho. The density uncertainty of Tycho (0.2-1.0~$\mathrm{cm^{-3}}$) is likewise reflected in the uncertainty in the distance to the remnant (see \cite{Hayato_2010} for a review and \cite{Tian_2011}), since both quantities are interdependent. For our modeling we choose a  value for the hydrogen number density of $n_\mathrm{H}=$0.6~$\mathrm{cm^{-3}}$, which gives a good fit to the observed shock radii and velocities.

\begin{figure}[tb!]
\includegraphics[width=0.49\textwidth, trim= 0.5mm 0.5mm 3mm 0.5mm, clip]{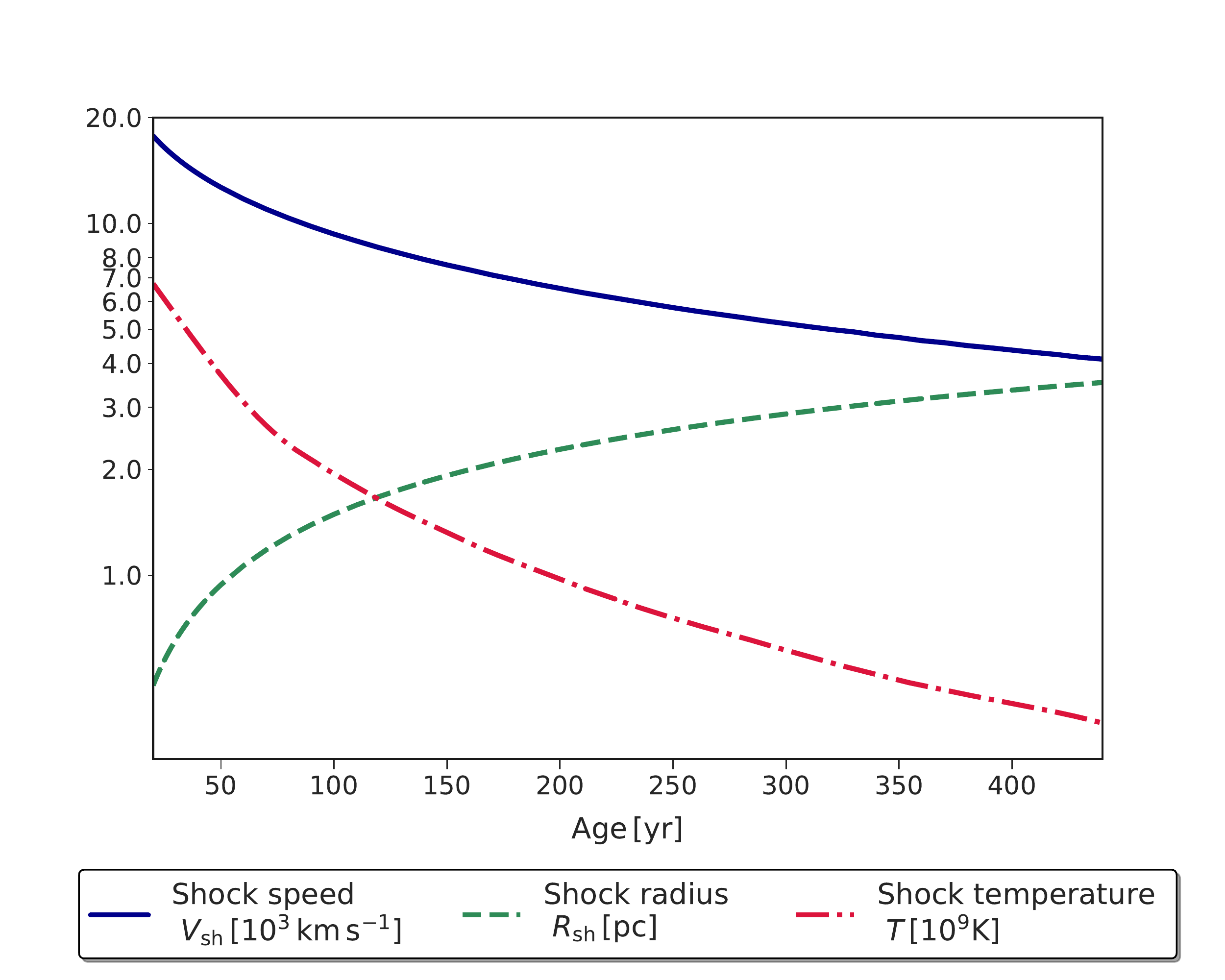}
\caption{The shock speed (blue solid line), radius (green dashed line) and temperature (red dotted line) as functions of time. }
\label{fig:hydro}
\end{figure}

Explosion energy, ejecta mass and ambient density, together with the exponential density profile, form the suite of parameters necessary to model the complete hydrodynamical evolution of Tycho. Our simulations are spherically symmetric, and are similar to those described in
\citet{dc98} and \citet{Teletal12a}. The
evolution of the remnant into the surrounding medium gives rise to a
double-shocked structure consisting of a forward shock expanding into
the surrounding medium, and a reverse shock that propagates back into
the ejecta. The two are separated by a contact discontinuity that divides
the shocked ejecta from the shocked surrounding medium.
The time-evolution of the shock speed, radius and temperature derived from our simulations is depicted
in Fig.~\ref{fig:hydro}.
At 440 years our simulations provide a forward shock radius $R_\mathrm{sh} \approx 3.5$~pc, which implies a distance to the remnant, $d\approx 2.9$~kpc.
The velocity of the forward shock yields $V_\mathrm{sh}\approx 4100\,\mathrm{km\, s^{-1}}$. The position of the reverse shock, $\sim0.69R_\mathrm{sh}$, is in good agreement with measurements of~\citet{2005ApJ...634..376W}. By contrast, the position of the contact discontinuity (CD), $R_\mathrm{CD}\approx 0.78 R_\mathrm{sh}$,
falls below the value identified by~\citet{2005ApJ...634..376W}, who interpreted the closeness of the CD to the blast wave as evidence for the efficient back-reaction of cosmic rays. However, global models that incorporate non-linear DSA (NLDSA) effects~\citep{2012A&A...538A..81M, 2014ApJ...783...33S} fail to reproduce the CD position for Tycho. The discrepancy for the CD position can be attributed
to the decelerating CD being unstable to the Rayleigh-Taylor instability. Two-dimensional hydrodynamical simulations with the exponential profile 
show that Rayleigh-Taylor structures can extend almost halfway from the CD to the outer shock \citep{vvd2000, 2001ApJ...549.1119W}. Furthermore, \citet{orlandoetal12} have shown that Rayleigh-Taylor instabilities and ejecta fingers that extend far beyond the CD can misleadingly suggest that the CD is further out than its actual location.
Therefore, we conclude that the position of the CD obtained in our model is quite reasonable.\\
According to our simulations, at the age of 440 years the remnant accumulated $\sim 3.8 M_{\odot}$ of ambient gas, indicating that in our modeling Tycho is in the transition between ejecta-dominated and Sedov-Taylor stages. The total thermal energy in the remnant at 440 years is $E_\mathrm{th}\approx 5.6\times 10^{50}$~erg.

The shock profiles and velocity distribution from the simulation are used in the calculation for the
particle acceleration, described in Section~\ref{sec:part_acc}.

\subsection{Magnetic field}
\label{sec:mag_fl}
In this work, the canonical value for the magnetic-field strength of the interstellar medium (ISM), $B_\mathrm{ISM}=5\, \mu\mathrm{G}$, in the far upstream is used. Its exact value is, however, insignificant for our modeling, since the magnetic field is amplified in the vicinity
of the shock, e.g. due to streaming instabilities \citep{BellsInst, Luc&Bell} or turbulent dynamos \citep{2007ApJ...663L..41G}.
Thus, the amplified magnetic field represents a physically relevant parameter. The exact treatment of the magnetic-field amplification is beyond the scope of this work. We adopt instead for the magnitude of the magnetic-field strength in the immediate upstream region $B_1=\alpha B_\mathrm{ISM}$, where $\alpha$, a generic amplification factor, is a free parameter in our model. Following the ansatz used by \citet{2008ApJ...678..939Z} and \citet{RB_MagTran} the magnetic-field profile in the precursor is assumed to drop exponentially to the interstellar field, $B_\mathrm{ISM}$, at the distance of 5\% of the shock:
\begin{equation}
B(r)=\begin{cases} \alpha B_\mathrm{ISM}\cdot \exp\left(\frac{-(r/R_\mathrm{sh}-1)}{0.05}\cdot\ln\left(\alpha\right)\right)& \text{for }  R_\mathrm{sh} \le r \le 1.05 \cdot R_\mathrm{sh}\\
B_\mathrm{ISM} &\text{for } r \ge 1.05 \cdot R_\mathrm{sh} \,.
\end{cases}
\end{equation} 
We stress that the magnetic-precursor length scale is not related to a free-escape boundary, which is usually introduced in the global modeling of SNRs~\citep{2008A&A...483..529V, 2012A&A...538A..81M, 2014ApJ...783...33S}. The precursor merely reflects typical characteristics of the spatial profile of the amplified magnetic field in the upstream region. In contrast to the previous models, our approach does not need an escape boundary, since our system is large enough (ca. 65 shock radii) to retain all injected particles in the simulation.
The pre-shock magnetic field is assumed to be isotropic, i.e its individual components are equal in their magnitudes (for more details see \cite{Teletal13}). Shock compression for the strong unmodified shock yields the immediate downstream value, $B_2 = \sqrt{11}\, B_1$. \\
We consider two scenarios for the magnetic-field distribution inside the SNR. In the first case,
the immediate downstream value of the magnetic field is transported inside the SNR with the plasma flow and evolves according to the induction equation for ideal magnetohydrodynamics (MHD) \citep{Teletal13}:

\begin{equation}
\frac{\partial \boldsymbol{B}}{\partial t}=\boldsymbol{\nabla} \times(\boldsymbol{u}\times \boldsymbol{B})\,.
 \label{eq:induc_eq}
\end{equation}
Here $\boldsymbol{B}$ is the magnetic-field vector and $\boldsymbol{u}$ the plasma velocity, obtained from hydrodynamic simulations. Eq.~\ref{eq:induc_eq} accounts for advection, stretching and compression of the field, implying that magnetic flux is conserved during the entire evolution of the SNR. 

For the second case, we assume that after being amplified in the upstream region and compressed at the shock, the magnetic field decays due to the damping of magnetic turbulence in the downstream region. 
Magnetic-field damping is one of the key processes in astrophysical plasmas \citep{Kulsrud.1971a, Threlfall.2011a}, and is expected to operate in SNRs, where it dissipates turbulently amplified magnetic field \citep{2005ApJ...626L.101P}. We do not know the turbulence mode that is most relevant for magnetic-field amplification in SNRs, and hence the particulars of damping are beyond the scope of this paper. Instead, we adopt a simple parametrization for the magnetic profile in the downstream region:
\begin{equation}
 B(r)=B_0+(B_2-B_0)\cdot \exp\left(\frac{-(R_\mathrm{sh}-r)}{l_\mathrm{d}}\right)\,,
 \label{Bdamp}
\end{equation}
where $l_\mathrm{d}$ is the characteristic damping length of magnetic turbulence. The residual field level, $B_0$, is the limit value, to which the magnetic-field tends in the far downstream, behind the CD and the reverse shock of the SNR. In contrast to the previous case, the magnetic flux is not conserved in the scenario described by Eq.~\ref{Bdamp}. Far inside the remnant the advection of magnetic field has to become important as the magnetic field approaches the constant value $B_0$. This inner field is, however, physically non-relevant. First of all, we do not include the acceleration of particles at the reverse shock, since we find the radiative contribution of those particles to be negligible. Furthermore, the bulk of the accelerated particles resides in the vicinity of the forward shock and hence do not experience the constant magnetic field, $B_0$. Most of the particles cannot cross the CD, which separates the swept-up material from the ejecta, since their dynamics is governed by advection.  Only a marginal fraction of the very-high energetic particles overcome the CD via diffusive transport. Their radiative contribution to the synchrotron spectrum does not exceed 1\% of the overall emission, and thus is negligible for the global SED modeling. Therefore, we can omit the advection of magnetic field for the second scenario that includes magnetic-field damping.\\  
The separate treatment of hydrodynamics and magnetic field does not allow us to include dynamic feedback from magnetic pressure. Therefore, we investigate the range where the impact of magnetic field on the shock structure remains negligible. We analytically solve the classical MHD-equations for a plane-perpendicular shock in steady-state, assume frozen-in plasma, and derive the dependency of the shock compression ratio on the magnetic-field strength. Apart from the magnetic field the resulting gas compression ratio depends mainly on shock speed and upstream density, which in our model are $V_\mathrm{sh}\approx$ 4100 {km} s$^{-1}$ (at 440 years) and $n_\mathrm{H}=0.6\,\mathrm{cm^{-1}}$ (see Sec.~\ref{sec:hyd_mf}). As the extensive magnetic-field pressure reduces the compression ratio of the shock, it consequentially softens the particle spectra. Hence, the corresponding particle spectral index deviates from the canonical DSA solution ($s=2.0$) by $\Delta s$, resulting in $s=2+\Delta s$. The resulting radio spectral index, $\sigma=0.5+\Delta \sigma$, is accordingly modified by the value $\Delta \sigma=\Delta s/2$. We choose $\Delta \sigma=0.01$ (and thus $\Delta s=0.02$) as the limit within which the magnetic-field pressure can be neglected, since the corresponding changes are hardly visible in the radio spectrum when compared to the observed data. We find for the sonic upstream Mach number of $M\approx3500$, which corresponds to the hydrodynamic quantities we use in this work for Tycho, that the magnetic field is dynamically unimportant for Alfv\'enic Mach numbers above $M_\mathrm{A,c}\approx 17$. For lower Alfv\'enic Mach numbers the deviation from the classical DSA-solution exceeds the values of $s=2.02$ and $\sigma=0.51$ and hence cannot be neglected. Taking into account that the magnetic-field pressure is exerted only by the tangential components, we find the magnetic-field limits $B_\mathrm{1,max}\approx120\,\mathrm{\mu G}$ for the upstream and $B_\mathrm{2,max}\approx400\,\mathrm{\mu G}$ for the downstream regions, respectively. Any higher magnetic field would efficiently lower the shock compression ratio and produce particle spectra with $s>2.02$ and radio spectra with $\sigma > 0.51$, and thus demand an MHD treatment.

\subsection{Particle acceleration}
\label{sec:part_acc}

To determine the evolution of the particle number density, $N(r,p,t)$, we numerically solve the {\em time-dependent} transport equation for cosmic rays

\begin{align}\label{eq:transport}
\nabla(&D_{r}\nabla N-\boldsymbol{u}N)-\frac{\partial}{\partial p}\left(N\dot{p}-\frac{\nabla \boldsymbol{u}}{3}Np\right)+\frac{\partial}{\partial p}\left(p^2D_{p}\frac{\partial}{\partial p}\frac{N}{p^2}\right)
\nonumber \\
&+Q=\frac{\partial N}{\partial t}\,.
\end{align}
Here $D_r= \xi\,c\,r_\mathrm{g}/3$ is the spatial diffusion coefficient, with the speed of light $c$ and the Larmor radius $r_\mathrm{g}$. The parameter $\xi$ is the ratio between the spatial diffusion coefficient and that for Bohm diffusion, sometimes referred as the gyrofactor. The Larmor radius and hence the spatial diffusion coefficient are calculated using the local magnetic-field strength. Since in our approach the magnetic field is spatially variable (see Sec.~\ref{sec:mag_fl}), the spatial diffusion coefficient also varies with position.
The synchrotron losses for electrons are included in Eq.~\ref{eq:transport} via $\dot p=(4e^4B^2)/(9c^6m^{4}_{e})p^2$, with the elementary charge $e$ and the electron rest mass $m_e$.
The plasma velocity, $\boldsymbol{u}$, is obtained from the hydrodynamic simulations.\\
The injection of particles is a complex issue, which is not fully understood. For simplicity, we use the thermal leakage injection model as presented in~\cite{2005MNRAS.361..907B}. The source term for electrons and protons is given by
\begin{equation}
 Q_{i}=\eta_\mathrm{i} n_{1,i} V_\mathrm{sh}\delta(r-R_\mathrm{sh})\delta(p-p_\mathrm{inj,i})\,,
\end{equation}
where index $i$ denotes the particle species (electrons and protons), $\eta_\mathrm{i}$ the corresponding injection efficiency and $n_{1,i}$ the particle number density in the upstream region.
The particles in our approach are mono-energetically injected. The associated injection momentum is multiple of their thermal momentum:
\begin{equation}
 p_\mathrm{inj,i}=4.45p_{\mathrm{th,i}}\equiv4.45\sqrt{2m_ik_\mathrm{B}T}\,,
\label{eq:inj_mom}
\end{equation}
where $k_\mathrm{B}$ is the Boltzmann constant and $T$ is the temperature of the plasma.

Electrons and ions are not in equilibrium at collisionless shocks. In fact, the current Rankine-Hugoniot ion temperature is about 30 keV, whereas the observed electron temperature in the post-shock medium is on the order of a keV after years of residence in the downstream region. The thermal leakage model is based on the requirement that particles see the shock as a sharp discontinuity and hence are injected only if their Larmor radius exceeds the width of the shock wave. This condition requires that electrons need to be pre-accelerated to around 100~MeV to participate in DSA. Particle-in-cell simulations provide evidence that thermal electrons can indeed be pre-accelerated by shock-surface and shock-drift acceleration~\citep{Matsumoto2017, Bohdan2017}. Tests demonstrate \citep{2019ApJ...874..119K} that this process yields an electron spectra shaped as a power low with a spectral index in the range $\sim(1.5-5.5)$. The exact value of the spectral index depends on the internal structure of the shock transition region. We approximate the two-step energization of electrons by DSA at all momenta and electron injection at very low momentum, which intends to replace an advanced, computationally far more expensive treatment of the pre-acceleration. Thus, the same particle temperature set for Eq.~\ref{eq:inj_mom} is supposed to mimic the pre-acceleration process of electrons.  Although, we do not know the exact value of the spectral index provided by the shock-drift acceleration, it only marginally affects our final spectra, as it was shown in~\citet{2015A&A...574A..43P}(Fig.~4).

Since the main goal of this work is to explain the broad SED of Tycho, we can neglect the thermal and supra-thermal protons. Indeed, as the hadronic $\gamma$-rays production starts from the threshold energy $\sim$0.3~GeV, the protons with momenta below $p_\mathrm{inj,p}$ are insignificant for the final photon spectrum. In contrast, the electrons at lower energies play an important role, as the corresponding synchrotron emission is visible in the radio range.

Besides the full time dependence of the particle acceleration and transport, the novel aspect in our work is the stochastic acceleration of particles presented in Eq.~\ref{eq:transport} by the momentum-diffusion coefficient $D_p$.
\citet{2015A&A...574A..43P} explicitly derived the momentum-diffusion coefficient for the fast-mode waves and demonstrated that for low-energy particles the acceleration time is independent of momentum and can be of the order of a few years. At higher energy
the process is expected to become less efficient. Stochastic acceleration is an important damping process of fast-mode
waves \citep[cf.][]{2014MNRAS.442.3010T}, and so it is efficient only in a thin region behind the forward shock with the thickness, $L_\mathrm{fm}$. A useful parametrization of the momentum diffusion coefficient, $D_p$, is then~\citep{2015A&A...574A..43P}

\be\label{eq:DiffCoeff}
D_p(r,p)=\begin{cases}
0 & \text{for } r<(R_\mathrm{sh}-L_\mathrm{fm})\\ 
\frac{p^2}{\tau_\mathrm{acc}} f(p) &\text{for } (R_\mathrm{sh}-L_\mathrm{fm}) \le r \le R_\mathrm{sh} \,.
\end{cases}
\ee
Here, $\tau_\mathrm{acc}$ denotes the acceleration time at small momenta, and $f(p)$ approximates 
the loss in acceleration efficiency at higher energies as a power law,
\be\label{eq:Correc}
 f(p)=\begin{cases}
1 & \text{for } p\le p_0\\ 
\left(\frac{p}{p_0}\right)^{-m} &\text{for } p \ge p_0 \,.
\end{cases}
\ee
The critical momentum $p_0$ and the power-low index $m$ are free parameters of our model for now.
It is important to note here that the energy in fast-mode waves has mostly a kinetic character. As the magnetic component of the fast-mode turbulence in the typical post-shock plasma is weak, it cannot amplify the magnetic field sufficiently. Therefore, in our approach other type of waves, such as streaming instabilities~\citep{BellsInst, Luc&Bell} or turbulent dynamos~\citep{2007ApJ...663L..41G} are required to provide the magnetic-field amplification at the shock, which is an important scale factor for scattering (see Sec.~\ref{sec:mag_fl}). There is not necessarily a simple scaling between momentum and spatial diffusion \citep[e.g.][Eq.14]{2004ApJ...614..757Y}. In addition, different types of turbulence may be responsible for the diffusive transport and stochastic re-acceleration \citep[e.g.][page 23]{2009ASSL..362.....S}. Therefore, the spatial and the momentum diffusion coefficients are independent in our model.

The energy in the fast-mode turbulence that occurs in the post-shock region can be primarily provided by the background plasma. To simplify matters, in our approach the energy density in the fast-mode waves scales with the thermal energy density of the post-shock background plasma 
\be
U_\mathrm{fm}=\epsilon U_\mathrm{th}\,.
\label{eq:U_fm}
\ee 
Here $\epsilon$ is the energy-conversion factor, which is assumed to be of the order of a few percent. The minor value of $\epsilon$ provides that the energy transfer from the main plasma flow can be neglected in our hydrodynamic simulations. For a strong shock expanding in a cold plasma the downstream thermal energy is approximately
\be
U_\mathrm{th}\approx \frac{9}{8}\rho_\mathrm{ISM} V_\mathrm{sh} ^2\,,
\label{eq:U_th}
\ee
where $\rho_\mathrm{ISM}$ denotes the ambient gas density.

The acceleration time for the fast-mode waves, derived in~\cite{2015A&A...574A..43P}, is given by
\begin{equation}
 \tau_\mathrm{acc}\approx (0.63\, \mathrm{yr})\left(\frac{U_{\textrm{th}}}{10\, U_{\textrm{fm}}}\right)\,.
\label{eq:tau}
\end{equation}
If energy density of the fast-mode waves scales with the thermal post-shock energy density of the plasma, as reflected by Eq.~\ref{eq:U_fm}, Eq.~\ref{eq:tau} simplifies to
\be
\tau_\mathrm{acc}\approx (6.3\,\mathrm{yr})\left(\frac{\epsilon}{0.01}\right)^{-1}\,,
\label{eq:tau_simpl}
\ee
providing that the acceleration time scale depends only on the energy fraction that is transferred to the fast-mode waves.

The re-acceleration of cosmic rays will inevitably damp the fast-mode waves and affect the width of the turbulent region, $L_\mathrm{fm}$.  The latter quantity can be estimated by equating the energy density of the fast-mode waves, $U_\mathrm{fm}$, and the total energy per volume that went into cosmic rays. Its value can be obtained from the energy transfer rate from waves to particles, $\dot{E}_\mathrm{tr}$,
and the time period that particles spend in the turbulent region interacting with fast-mode waves, $\Delta t$, providing
\be
  U_\mathrm{fm}=\dot{E}_\mathrm{tr}\cdot\Delta t = \frac{U_\mathrm{cr}}{\tau_\mathrm{acc}}\cdot\frac{L_\mathrm{fm}}{u_2}\,.
\label{eq:cr_energy}
\ee

Here $U_\mathrm{cr}$ is the cosmic-ray energy density in the immediate post-shock region induced by the stochastic re-acceleration. Combining Eqs.~\ref{eq:tau} and \ref{eq:cr_energy} we finally obtain
\begin{equation}
L_\mathrm{fm}\approx(5\times 10^{13}\, \mathrm{cm}) \left(\frac{V_\mathrm{sh}}{1000\, \mathrm{km/s}}\right)\left(\frac{U_\mathrm{th}}{U_\mathrm{cr}}\right)\,.
\label{eq:turb_reg}
\end{equation}
Summarizing, the thickness of the re-acceleration region is limited by damping of the turbulence caused by the re-acceleration of particles.
Therefore, in our method $U_\mathrm{cr}$ from Eq.~\ref{eq:turb_reg} is coupled to the intermediate results from Eq.~\ref{eq:transport} for the immediate post-shock area.
Calculating the energy density of cosmic rays, $U_\mathrm{cr}$, we do not take the contribution from electrons into account, since in our model their energy is negligible compared to that of protons.

The physically essential quantity for stochastic re-acceleration is the amount of energy available for it, here in the form of fast-mode waves and described by the parameter $\epsilon$. The size of the turbulence region, $L_\mathrm{fm}$, follows from the acceleration time scale, $\tau_\mathrm{acc}$, and only their ratio is relevant for the resulting particle spectra. The reason is that re-acceleration, once efficient, becomes the main damping mechanism for the waves and hence ceases when the energy supply is exhausted. More details on general discussion on stochastic re-acceleration in SNRs can be found in \cite{2015A&A...574A..43P}. 

\subsection{Radiative processes}\label{rad_pr}
To account for the entire SED from Tycho, we calculate synchrotron radiation of electrons and $\gamma$-ray radiation resulting from leptonic and hadronic interactions. The synchrotron emission is calculated following \citet{BlumGould:1970}.
Care must be exercised to properly account for magnetic field fluctuations in calculating the synchrotron emissivity function. The standard calculation, although not applicable
to the turbulent magnetic field, assumes by default a delta function for the probability distribution of local magnetic-field amplitudes. For the turbulent magnetic-field the amplitudes distribution is dispersed compared to the standard case. Its exact form is unknown. In the literature, however, one finds modified emissivities for exponential \citep{2010ApJ...708..965Z} and power-law \citep{2013ApJ...774...61K} distributions.
In this work we use the Gaussian distribution of magnetic-field amplitudes and the corresponding synchrotron emissivity function \citep{2015A&A...574A..43P}.
Given the low gas density, non-thermal bremsstrahlung \citep{BlumGould:1970} yields too low a flux for it to be relevant, and so leptonic gamma-ray emission is solely provided by inverse Compton scattering, for which we consider the microwave background as the target photon field \citep{BlumGould:1970, Sturner:1997}.
Tycho shows evidence for infrared emission \citep{2001A&A...373..281D}. Its contribution for the overall inverse-Compton spectrum,
investigated by \cite{2011ApJ...730L..20A} was, however, found to be negligible. 
Hadronic $\gamma$-rays are the result of decays of neutral pions and other secondaries produced in interactions of cosmic rays with nuclei of the ISM. To calculate its spectrum we follow the method from \citet{Huang:2006bp}.

\section{Results}
The method to solve the transport equation (Eq.~\ref{eq:transport}) separately for electrons and protons is described in \citet{Teletal12a}. We emphasize that, in contrast to previous attempts to model Tycho, we follow the full temporal evolution of the remnant starting at an age of 25 years. For conciseness, however, we show and discuss only results for the current age of Tycho. We find that the reverse-shock contribution to the particle spectrum in Tycho is negligible in the framework of our modeling at an age of 440 years \citep{Teletal12a}, in full agreement with \cite{2005ApJ...634..376W}. Therefore we do not discuss it further.

We present two models for the multiwavelength emission of Tycho, which both adequately fit the SED.
Model~I allows for the weakest magnetic field in the immediate downstream region of the shock, $B_2 = 150\, \mathrm{\mu G}$, that is compatible with the entire $\gamma$-ray flux observed with \textit{Fermi}-LAT in GeV-range and VERITAS in TeV-band \citep{gamma_data}. In Model~I, the resulting $\gamma$-ray flux consists of both leptonic and hadronic components. The magnetic
profile deeper inside the remnant is determined by advection of frozen-in magnetic field and corresponds to the MHD solution for
negligible magnetic pressure and energy density. As we shall demonstrate, this model fails to explain radio and X-ray intensity profiles.

Therefore we present a second model, Model~II, involving magnetic field damping and derive constraints on the magnetic field in Tycho.
Generic technical details can be found in subsection~\ref{moda}, where we introduce Model~I. They also apply to the following subsections listing the results for Model~II.

\subsection{\emph{Model~I:} Moderate advected magnetic field}\label{moda}

\begin{table*}[!ptb]
\centering
 \caption{Summary of the model parameters.}
\begin{threeparttable}
 \begin{tabular}{ c c c c c c | c c c c c}
 \hline
&\multicolumn{5}{c|}{ Varying parameters} &\multicolumn{5}{|c}{Fixed for both models}\\
\hline 
 Model  &  $B_2$ & $B_0$& $l_\mathrm{d}$  &$\xi$   & $\eta_\mathrm{e}$ &$\eta_\mathrm{p}$& $n_\mathrm{H}$& $p_0$ & $m$&$\epsilon$\\
 & ($\mathrm{ \mu G}$)& ($\mathrm{ \mu G}$) &($R_\mathrm{sh})$ & & & & ($\mathrm{cm}^{-3}$)&  ($10^{-4}m_\mathrm{p}c$)& &\\
 \hline
 I  & 150 & -& - & 10 & $9.4\times10^{-6}$& $2.4\times10^{-5}$ & 0.6 & 9.3 &0.25& 0.027 \\
 II & 330& 70 & 0.01 &16  & $10.2\times10^{-6}$ & $2.4\times10^{-5}$& 0.6& 9.3 & 0.25& 0.027 \\
 \hline
 \end{tabular}
\vspace{4mm}
{\tiny
 \begin{tabular}{ l l c l l}

  $B_\mathrm{2}$ & magnetic field in the immediate post-shock region &
 &$\eta_\mathrm{p}$& injection efficiency of protons\\
  $B_0$ & residual level of the magnetic field &
&  $n_\mathrm{H}$& ambient hydrogen number density \\
  $l_\mathrm{d}$ &damping scale of the downstream magnetic field &
 &$p_0$& critical momentum of the momentum-diffusion coefficient\\
  $\xi$ &spatial diffusion coefficient parameter (gyrofactor) &
 & $m$& power-low index of the momentum-diffusion coefficient\\
 $\eta_\mathrm{e}$ & injection efficiency of electrons&
& $\epsilon$ &energy-conversion factor for turbulence

 \end{tabular}
}

\end{threeparttable}
 \label{Theor_param}
\end{table*}
The efficacy of stochastic acceleration is fully determined by three parameters: $m$ and $p_0$, and the energy fraction of the plasma transferred to the turbulence, $\epsilon$. The acceleration time scale, $\tau_\mathrm{acc}$, and the width of the turbulent region, $L_\mathrm{fm}$, are then self-consistently calculated according to Eqs.~\ref{eq:tau_simpl} and~\ref{eq:turb_reg}. The parameter $p_0$ provides the critical limit up to which the momentum diffusion coefficient $D_p$ is energy-independent (cf. Eq.~\ref{eq:Correc}). It should be derived from exact scattering theory, which is, however, beyond the scope of this work. We tested several values of $p_0$ and found that for $p_0\gtrsim10^{-3}\,m_\mathrm{p}c$ the resulting curvature of the radio synchrotron spectrum becomes too strong to remain in agreement with radio data. Therefore, in this work we adopt $p_0=9.3\times10^{-4}m_\mathrm{p}c$, which corresponds to an electron energy of 500 keV. The power-law index, $m$, may vary up to 25\% (to stay in agreement with observations) and can be compensated by suitable choice of the energy-conversion factor, $\epsilon$, implying that these quantities exhibit degeneracy to a certain degree. In the following calculations we will fix $m = 0.25$ and $\epsilon=0.027$ (corresponding to 2.7\% of the thermal energy density), which adequately reproduce the observed radio spectrum. The acceleration time scale is determined by the energy-conversion factor, which for $\epsilon=0.027$ provides $\tau_\mathrm{acc}\approx2.3$~yr.

\begin{figure}[htb!]
\includegraphics[width=0.49\textwidth, trim= 0.5mm 0.5mm 3mm 0.5mm, clip]{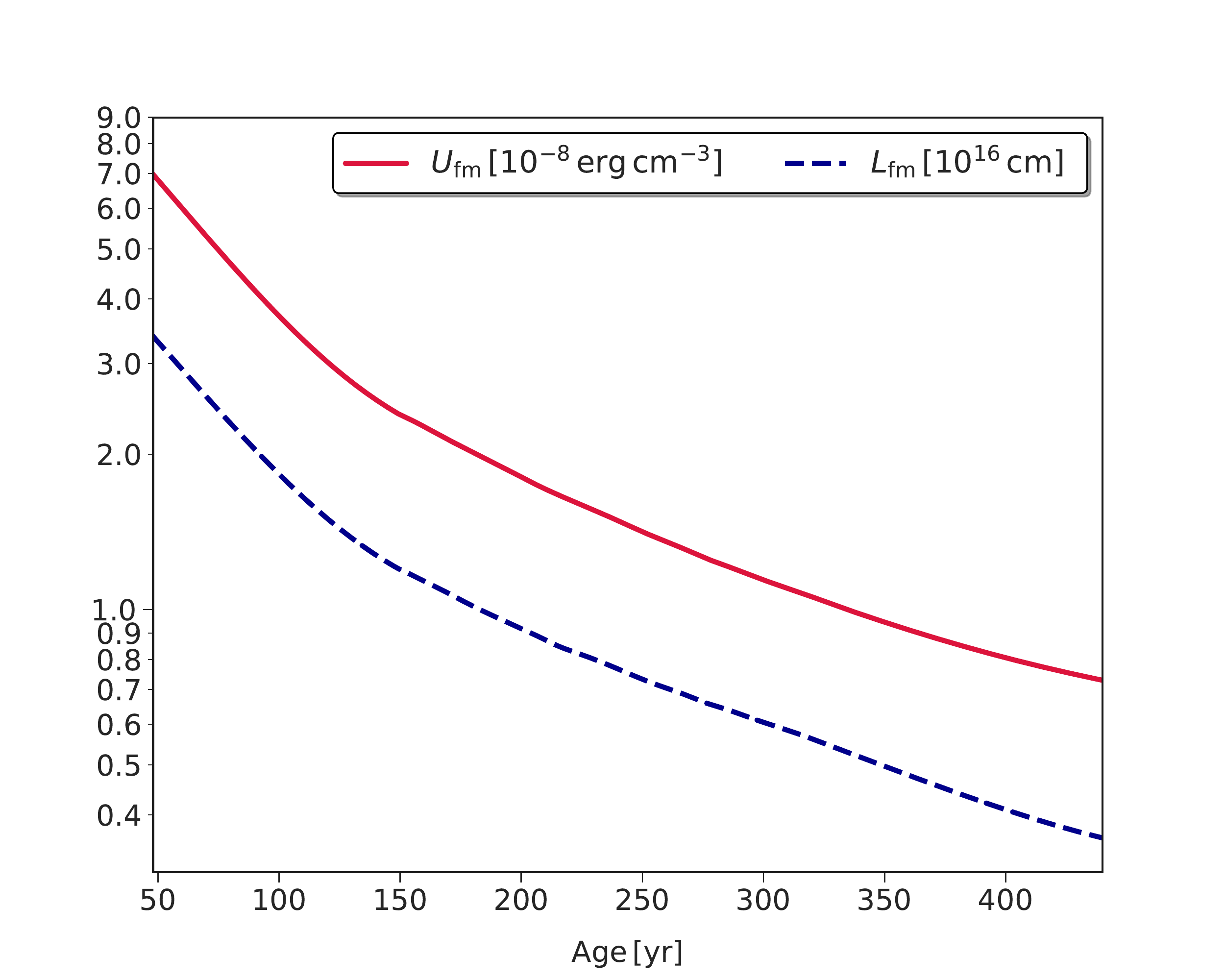}
\caption{The energy density in fast-mode waves (red solid line) and width of the turbulent re-acceleration region (blue dashed line) as functions of time for Model~I.}
\label{fig:Ufm_n_l_tur}
\end{figure}

The thickness of the region, in which stochastic re-acceleration is operating, is fairly small. For 2.7\% of the thermal energy density of the post-shock plasma transferred to the turbulence, it results into $L_\mathrm{fm} \sim 10^{15}-10^{16}\,\mathrm{cm}$, as illustrated in Fig.~\ref{fig:Ufm_n_l_tur}. The width of the re-acceleration region decays with time as the SNR shock slows down and thus the energy density in the turbulence decreases. At the age of 440 years the width of the turbulent region comprises only $\sim3\times10^{-4}R_\mathrm{sh}$. Still, the contribution to the particle spectra is substantial.

\begin{figure}[tb!]
\includegraphics[width=0.49\textwidth, trim= 0.5mm 0.5mm 3mm 0.5mm, clip]{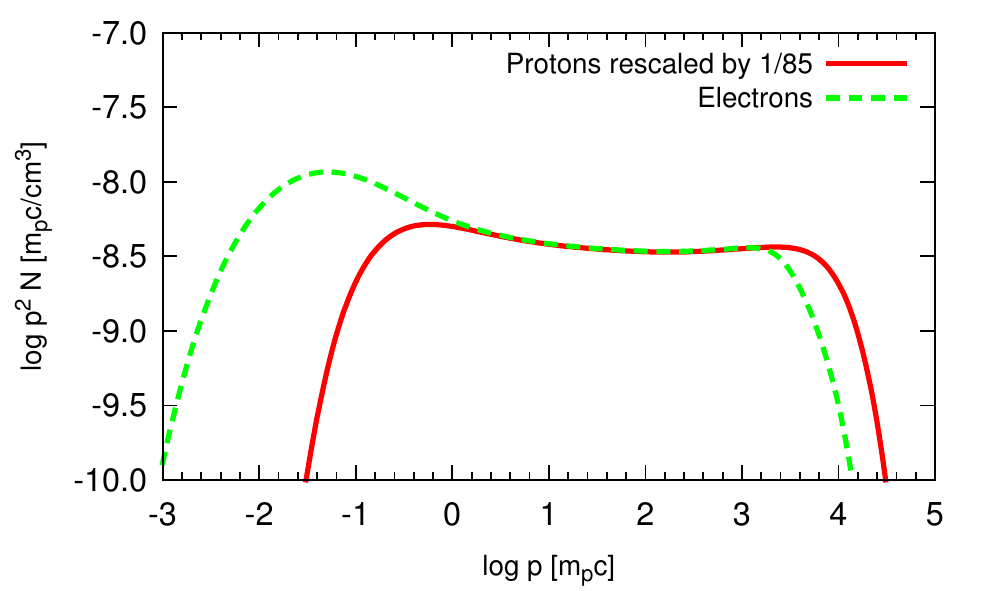}
\caption{The number density of protons (red solid) and electrons (green dashed) for Model~I involving weakest advected magnetic field with $B_2 = 150\,\mu$G. The proton number density is multiplied with the factor $K_\mathrm{e/p} = 1/85$. }
\label{fig:part_spec_mom}
\end{figure}

\begin{figure}[tb!]
\includegraphics[width=0.49\textwidth, trim= 0.5mm 0.5mm 3mm 0.5mm, clip]{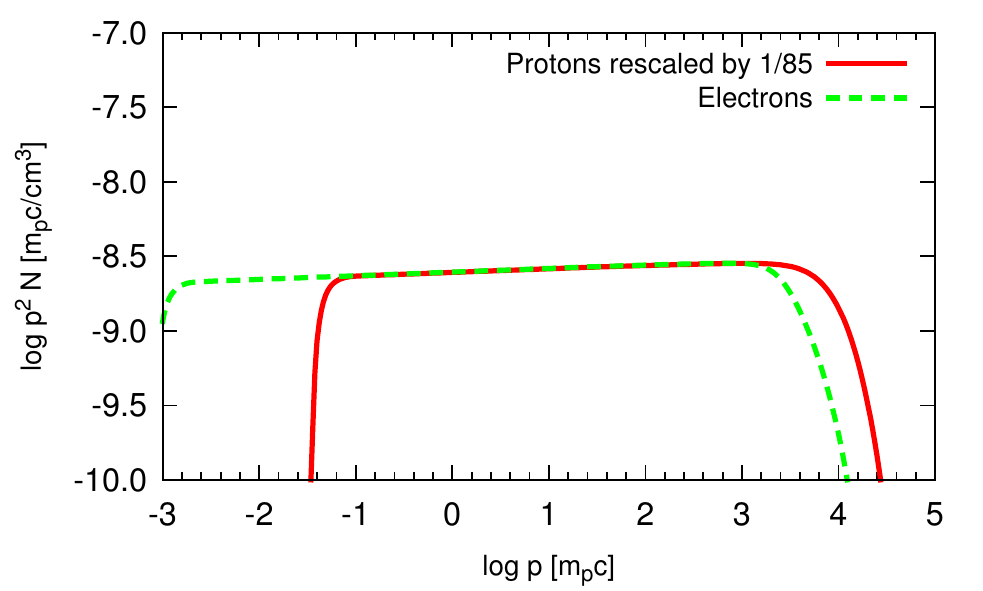}
\caption{The number density of protons (red solid) and electrons (green dashed) for Model~I with turned off stochastic re-acceleration. A comparison with Fig.~\ref{fig:part_spec_mom} illustrates the quantitative effect of the second order acceleration in cosmic-ray spectra.}
\label{fig:part_spec_mom_nosofa}
\end{figure}

Fig.~\ref{fig:part_spec_mom} shows the differential number density for electrons and protons as obtained from solving Eq.~\ref{eq:transport}.
As a result of stochastic re-acceleration, the spectra strongly deviate from the canonical solution, $N \propto p^{-2}$, expected for DSA in the test-particle limit. For comparison, a standard case with stochastic re-acceleration turned off but the same remaining parameters is depicted in Fig.~\ref{fig:part_spec_mom_nosofa}.
To be noted in Fig.~\ref{fig:part_spec_mom} are distinct bumps for both particle species at low energies with concave tails that extend up to the maximum momenta of the spectra, $\sim10^3m_pc$.
Deviations between electron and proton spectra at lower momenta results from the different injection criteria for the particles. Since we use the thermal leakage model, electrons and protons feature injection momenta differing by a factor $\sim 40$ due to $p_\mathrm{inj,i}\propto\sqrt{m_i}$, as seen from Eq.~\ref{eq:inj_mom}. In other words, we do not explicitly treat electron acceleration by, e.g., stochastic shock drift acceleration below 100~MeV \citep{2019ApJ...874..119K} and subsume the entire acceleration from suprathermal to very high energies under DSA, as mentioned above. Electrons and protons at energies between the thermal peak and the injection threshold a factor 4.45 higher in momentum are considered part of the thermal bulk and ignored. The electrons between their injection threshold and the proton injection momentum are very numerous and with stochastic re-acceleration can form a larger bump in the particle spectrum than would be observed for protons. 

The total energy that went into electrons, $E_\mathrm{tot, e}\approx 6.8\times10^{47}\,\mathrm{erg}$, is marginal compared to that of protons, $E_\mathrm{tot, p}\approx 2.7\times10^{49}\,\mathrm{erg}$. Reasons for this are the small rest mass of electrons and different injection efficiencies $\eta_e=9.4\times10^{-6}$ and $\eta_p=2.4\times10^{-5}$, which are chosen to fit the entire SED. The electron injection efficiency simultaneously accounts for the maximum IC peak consistent with the $\gamma$-ray data and sufficient radio emission for $B_\mathrm{d}=150\,\mathrm{\mu G}$ (amplification factor $\alpha\approx 9$), while injection of protons provides an adequate hadronic contribution in the GeV range. The injection efficiencies determine the electron-to-proton flux ratio, $K_\mathrm{e/p} \equiv {N_e}/{N_p} \approx {1}/{85}$ in the range $(10-10^{3})m_p\,c$.

\begin{figure}[tb!]
\includegraphics[width=0.49\textwidth, trim= 0.5mm 0.5mm 3mm 0.5mm, clip]{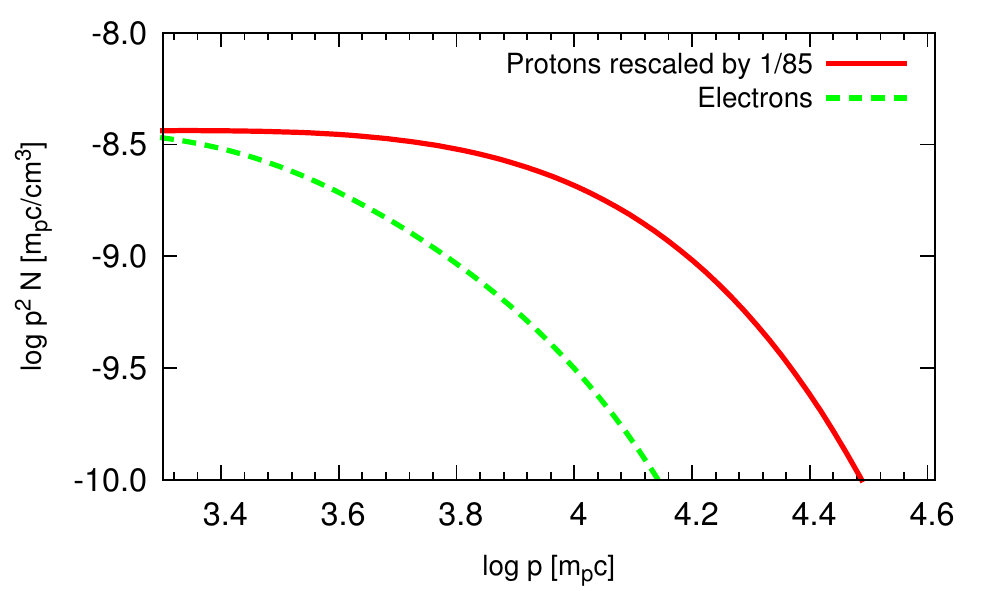}
\caption{The number density of protons (red solid) and electrons (green dashed) for Model~I as in Fig.~\ref{fig:part_spec_mom} zoomed to the cutoff region.}
\label{fig:part_spec_mom_cutoff}
\end{figure}
The proton spectrum cuts off at the maximum momentum, $p_\mathrm{max}\approx10^4 m_p\,c\simeq 10\,\mathrm{TeV/c}$, which is limited by the age of the remnant and particle diffusion in the upstream region. The latter process is determined by two factors: the magnetic precursor which flattens the cosmic-ray precursor, discussed in Sec.~\ref{sec:mag_fl}, and the gyrofactor, $\xi$, which beside the magnetic-field strength co-determines the diffusive transport of particles.
Since the electrons additionally experience effective synchrotron losses, their maximum momentum, $\sim 3\,\mathrm{TeV/c}$, is lower than the limit  ($10\,\mathrm{TeV/c}$) imposed by the spatial diffusion and the age of the remnant. The exact shape of electron and proton cutoffs is shown in Fig.~\ref{fig:part_spec_mom_cutoff}, which allows a direct comparison of the cutoff forms. As one can see, the electron density is steeper than that of protons due to effective synchrotron losses. Note that the shape of the spectral cutoff in our approach deviates from the simple exponential and super-exponential forms (e.g. \cite{2010MNRAS.402.2807B}) due to the full time-dependency of our method. In contrast to the previous works on Tycho, our approach includes time-dependent transport equation and hydrodynamics. In fact, the maximum cosmic-ray energy in the age-limited case scales linearly with time, and as the square of the shock velocity, $V_\mathrm{sh}^2$. As the shock of the remnant slows down in our simulations, the maximum energy of cosmic rays grows more slowly than one would expect for an approach with a constant shock velocity. As a result, the final particle spectrum in our work, which is qualitatively an average of the instantaneous spectra, gives a sharper cutoff than predicted within the hydrodynamics in steady-state.

\begin{figure}[tb!]
\includegraphics[width=0.49\textwidth, trim= 2.5mm 0.5mm 3mm 0.5mm, clip]{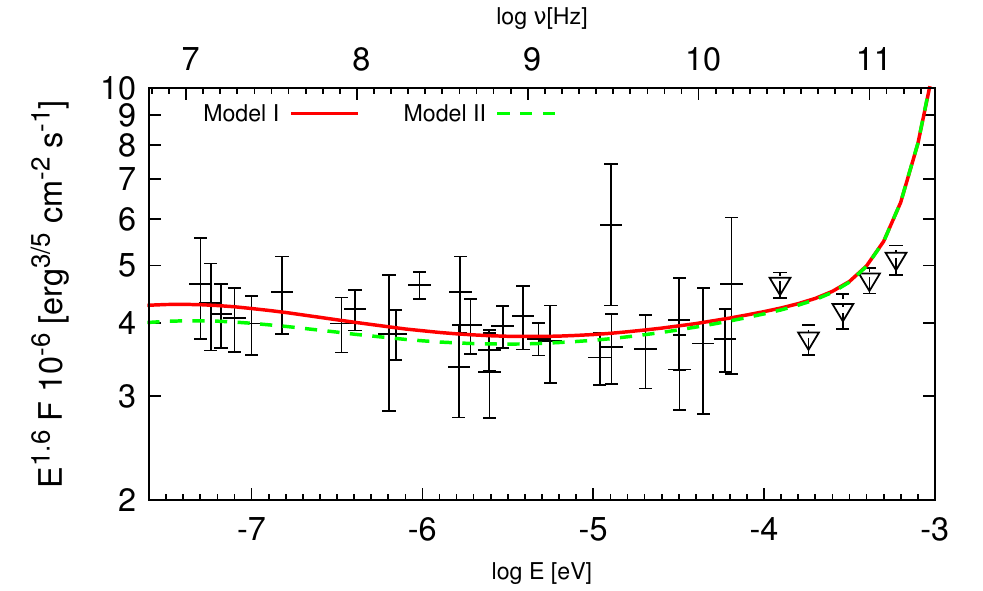}
\caption{Synchrotron plus gray-body emission in the radio and microwave band for Models I and II. The radio (black error bars) and microwave (black triangles) data are taken from \citet{1992ApJ...399L..75R} and \citet{Planck2016}. }
\label{fig:radio_planck1}
\end{figure}

Instead of showing the entire SED at one singe plot, we present figures for particular energy bands (radio, X-ray and $\gamma$-ray) to provide a detailed comparison of data and models. The synchrotron emission from electrons in the radio and microwave band for Model~I is presented as red solid line in Fig.~\ref{fig:radio_planck1}, where the data are taken from \citet{1992ApJ...399L..75R} and \citet{Planck2016}. Tycho's radio spectrum is frequently considered to be distorted in response to shock modification by cosmic rays \citep{1992ApJ...399L..75R,2008A&A...483..529V}. Fig.~\ref{fig:radio_planck1} clearly demonstrates that this is not the only viable interpretation: stochastic re-acceleration in the downstream region of a test-particle shock can reproduce the observed radio spectrum without invoking non-linear effects.\footnote{Shock modification by cosmic-ray feedback is neglected in our approach since the cosmic-ray pressure remains well below 10$\%$ of the shock ram pressure \citep{2010ApJ...721..886K}. In fact it is at most 2.1\% of the shock ram pressure (cf. Sec.~\ref{sec:crp}).}
It was noted earlier \citep{Planck2016} that the radiation from Tycho in the microwave band consists of at least two components: synchrotron emission from the non-thermal electrons and thermal dust emission. The latter process is responsible for the sharp rise in flux above $\sim$30~GHz. We account for the thermal dust emission by calculating the gray-body radiation with a temperature of 25~K and a normalization chosen to fit the flux density measured with Herschel \citep{2012MNRAS.420.3557G}. Thus, the red line in Fig.~\ref{fig:radio_planck1} represents the sum of the synchrotron and thermal spectra. While the slightly concave part of the radio spectrum below $\sim$30~GHz is dominated by the synchrotron emission from electron population shaped by the stochastic re-acceleration, the steep flux increase above $\sim$30~GHz results from the thermal dust emission.

\begin{figure}[tb!]
\includegraphics[width=0.49\textwidth, trim= 2.5mm 1mm 4mm 2mm, clip]{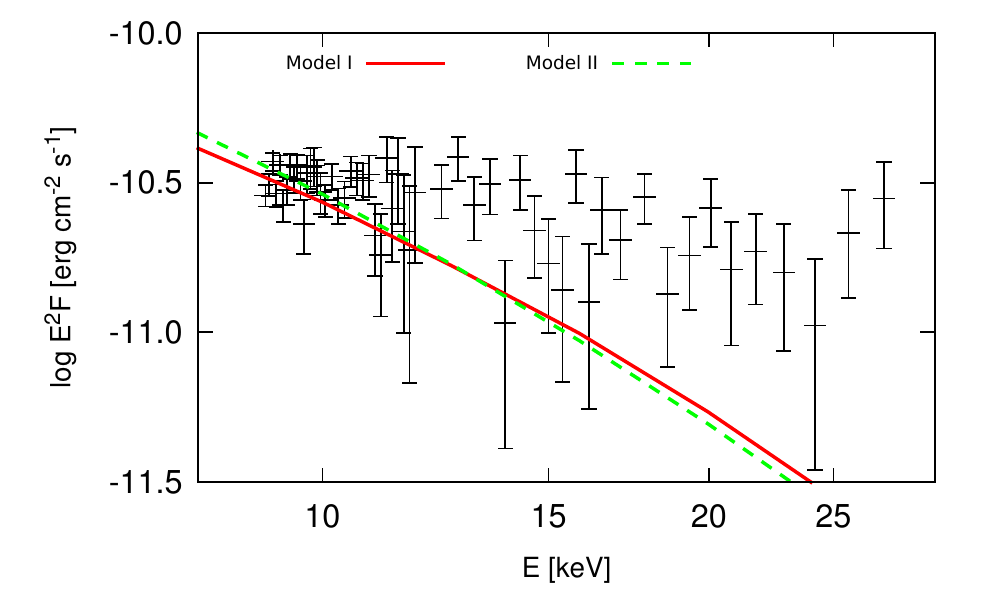}
\caption{Synchrotron emission in the X-ray band for Models I and II with Gaussian distribution of magnetic-field amplitudes. Experimental data at 90\% confidence level are taken from \citet{2009PASJ...61S.167T}.}
\label{fig:xray_emiss}
\end{figure}
The X-ray emission is generated by electrons beyond the cutoff energy, $E_\mathrm{max,e}\approx$ 3 TeV, via synchrotron radiation.
It is presented for Model~I as red solid line in Fig.~\ref{fig:xray_emiss}, where the experimental data at 90\% confidence level are taken from \citet{2009PASJ...61S.167T}. To achieve a good fit for the fixed magnetic field ($B_2=150\, \mathrm{\mu G}$) and electron injection efficiency ($\eta_e=9.4\times10^{-6}$) we adapt the spatial diffusion coefficient parameter and set $\xi=10$.
As already discussed in Section~\ref{rad_pr}, for our calculation of the synchrotron emission we use a Gaussian distribution function for the amplitudes of the turbulent magnetic field. The differences to the standard formula are mostly seen at the cutoff of
the synchrotron emission. Indeed, the standard emissivity function produces the steepest slope in the spectral tail while any extended distribution smears the spectral cutoff and causes spectral hardening in the X-ray band. Nevertheless, the predicted X-ray spectrum above 10 keV is softer than that observed. This can be attributed to the spherical symmetry of our model geometry. \cite{2015ApJ...814..132L} analyzed sixty-six regions across Tycho and find that their X-ray emission exhibits a varying roll-off energy, which is defined as

\begin{equation}
E_\mathrm{rolloff}\simeq 7\,\mathrm{eV}\left(\frac{B_2}{100\,\mu\mathrm{G}}\right)
\,\left(\frac{E_\mathrm{max, e}}{\mathrm{TeV}}\right)^{2}\,.
\label{eq:rolloff}
\end{equation}
As the magnetic field and the maximum energy of electrons may vary across Tycho's perimeter, so do the corresponding synchrotron spectra. Integration over the individual regions, as in \cite{2015ApJ...814..132L}, results in a harder total spectrum, because the variations in the roll-off energy will invariably smear out the cutoff. This hardening can not be accounted for in our model due to the 1-D geometry, and hence our X-ray spectrum is softer than that observed by \textit{Suzaku}. Note, since we use a Gaussian distribution function to calculate the synchrotron emission, its cutoff deviates from the usually assumed exponential profile and thus Eq.~\ref{eq:rolloff} is no longer valid in our case.

\begin{figure}[tb!]
\includegraphics[width=0.49\textwidth]{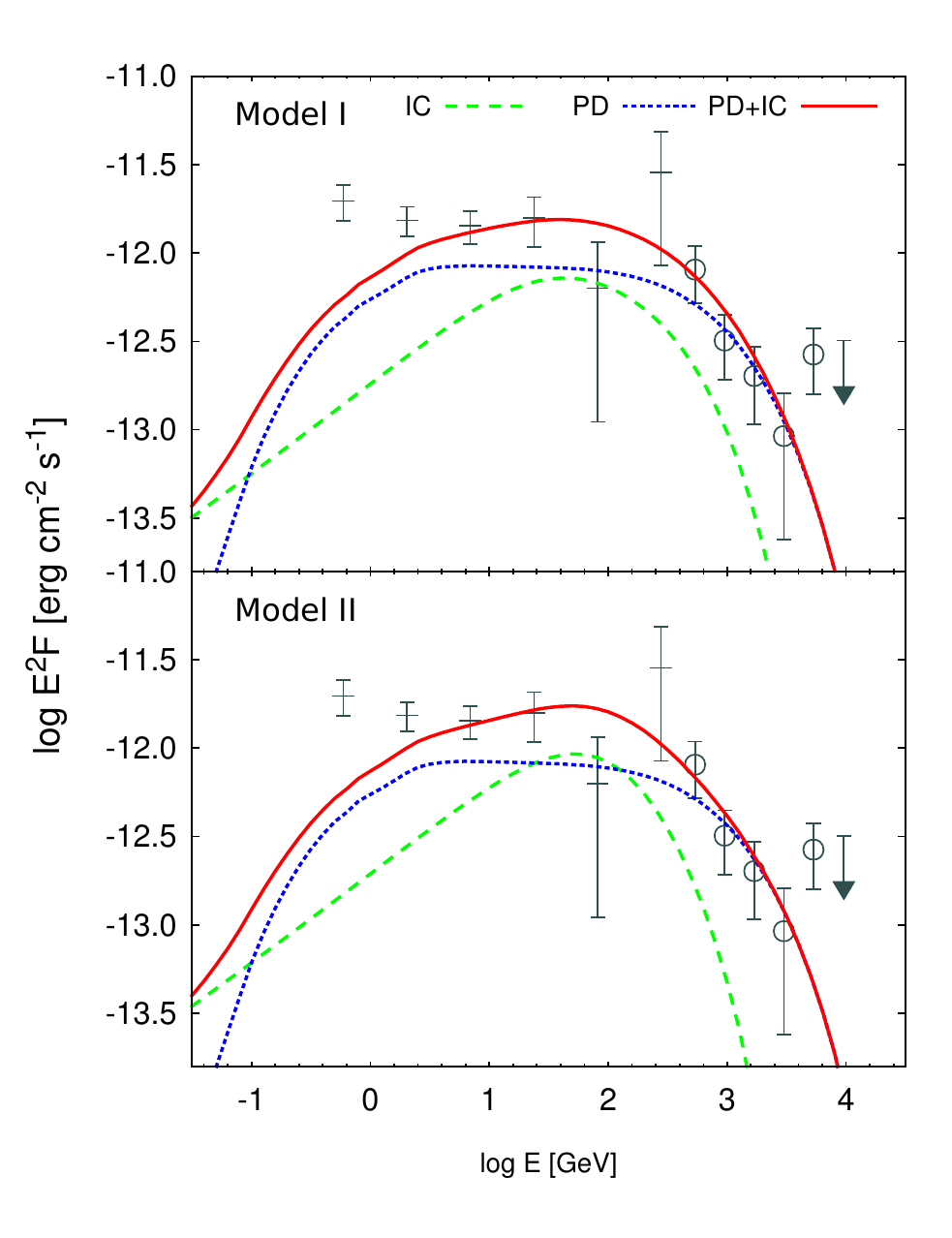}
\caption{The calculated $\gamma$-ray emission from Tycho in comparison
with measurements of \textit{Fermi}-LAT (black solid) and VERITAS (black circles) from \cite{gamma_data}.
The top panel is for Model~I involving a moderate transported magnetic field and the bottom panel presents results for Model~II, which includes damping of magnetic field.}
\label{fig:gamma1}
\end{figure}

The minimum downstream magnetic field that provides the maximum inverse-Compton contribution compatible with the $\gamma$-ray data and sufficient flux contribution in the radio range on account of stochastic re-acceleration is $\sim 150\,\mathrm{\mu G}$. Any weaker magnetic field would imply an overshooting of $\gamma$-ray emission at $\sim 100\,\mathrm{GeV}$ induced by the inverse Compton process.
Nevertheless, it is not able to account for the GeV-scale emission and hence the hadronic channel is required. The resulting $\gamma$-ray spectra of Tycho and the corresponding $\gamma$-ray data are given at the top panel of Fig.~\ref{fig:gamma1}. The pion bump is represented by the blue dotted and the inverse-Compton peak by the green dashed lines, respectively.
The total $\gamma$-ray spectral distribution (red solid line) is rather flat with spectral index
$\Gamma \approx 2$. The impact from the stochastic re-acceleration of protons is hardly visible in the pion bump. The reason is a relatively small energy fraction in the fast-mode waves. In our model for Tycho parameters relevant for the stochastic re-acceleration are dictated by the radio data. For other SNRs their value can differ, potentially resulting in more efficient re-acceleration of protons and consequentially into a softer hadronic spectrum.
Thus, stochastic acceleration of protons may provide an alternative explanation for the softening of hadronic emission, as opposed to high-density structures in the ambient medium~\citep{2013ApJ...763...14B, MorBla_NH}, if enough energy is available for re-acceleration. Nevertheless, the impact from stochastic re-acceleration is more prominent for electron than proton spectra. The cutoff in the leptonic and hadronic $\gamma$-ray contributions is linked to that in the synchrotron spectrum via the gyrofactor, $\xi$, and hence constrained by X-ray data. Thus, in our approach, we do not specifically adjust the model parameters to fit the hadronic cutoff. Nevertheless, it shows a remarkable match with the observed VERITAS data. The corresponding parameters for Model~I are summarized in the first row of Table~\ref{Theor_param}. The amplification efficiency (i.e. the ratio between cosmic-ray pressure and the amplified magnetic-field pressure in the immediate upstream region) is roughly $\sim 74$ for 440 years.

As already mentioned in Sec~\ref{rad_pr}, the contribution from non-thermal bremsstrahlung is negligible for the density of gas at Tycho. Therefore it is not shown in Fig.~\ref{fig:gamma1} and we do not discuss it further.

\begin{figure}[bt!]
\includegraphics[width=0.49\textwidth]{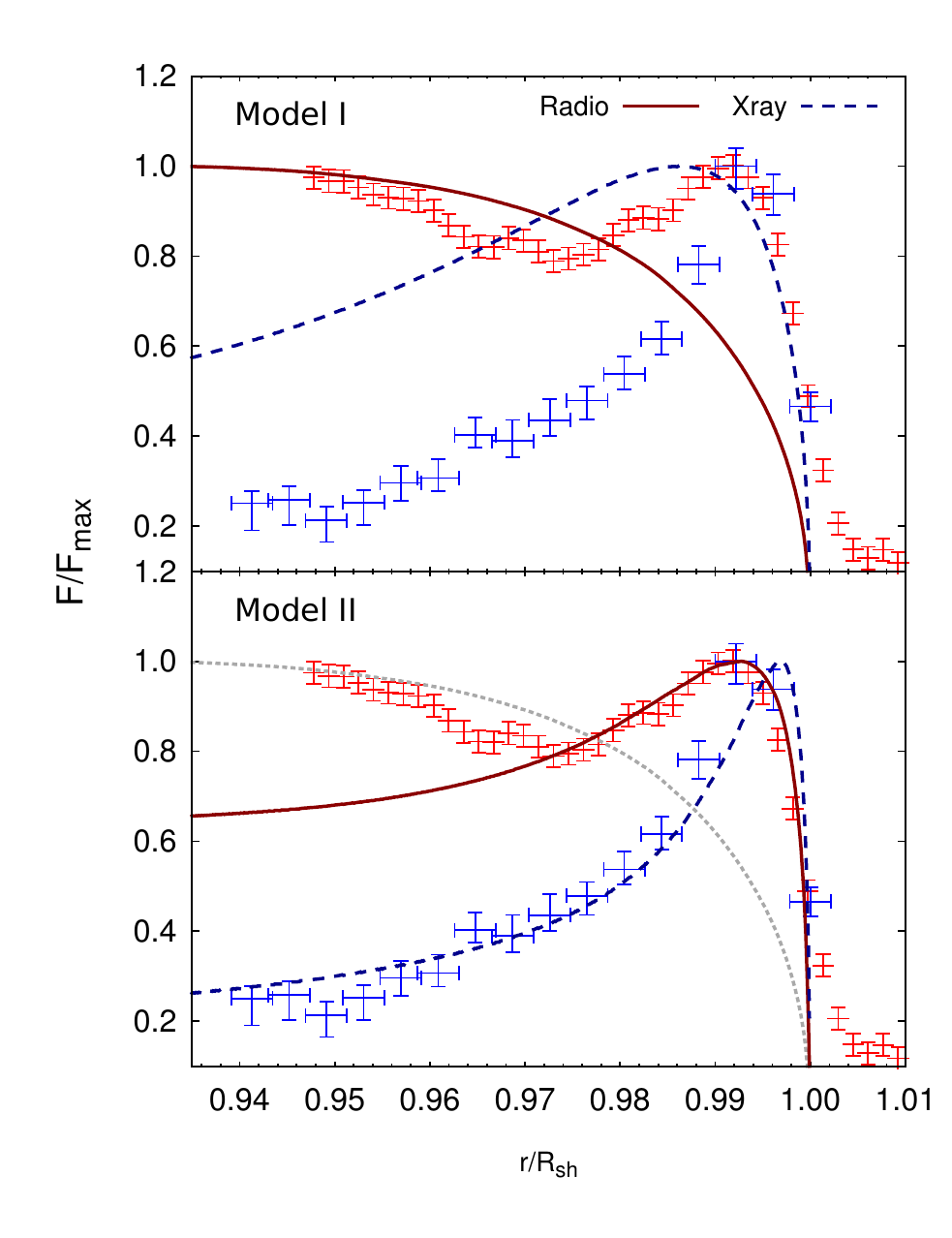}
\caption{Radial intensity profiles for Model~I (top panel) and Model~II (bottom panel). X-ray emission at 1 keV (blue dashed line) was observed with \textit{Chandra} \citep{bw_Cas_Chen}, and radio data at 1.4~GHz (red solid line) were taken from \cite{2014ApJ...783...33S}.
Following \cite{2014ApJ...783...33S} the radio data were slightly shifted to account for the expansion of the remnant. The grey dotted line in the bottom panel represents a radio profile produced without magnetic-field damping.}
\label{fig:xrprof1}
\end{figure}

To further test Model~I we consider the spatial profiles of radio and X-ray synchrotron radiation at 1.4~GHz and 1~keV, respectively. The results are depicted in top panel of Fig.~\ref{fig:xrprof1}. The radio \citep{2014ApJ...783...33S} and X-ray \citep{bw_Cas_Chen} brightness profiles are extracted from the western rim of the remnant and normalized to their peak values.
The measured X-ray brightness is strongly peaked at the shock and rapidly decreases towards the contact discontinuity. The radio profiles are slightly wider but still exhibit a narrow structure close to the shock. The enhancement of the emission towards the interior that is evident in the radio profiles might be attributed to the afore-mentioned Rayleigh-Taylor distortions operating in the vicinity of CD. As seen in the figure, the predictions of Model~I involving the smallest possible advected magnetic field are in disagreement with the data.
The model can explain neither the narrow X-ray rims nor the structure of the radio emission. 

There are at least two potential solutions to this problem that we shall explore in the following section. Turbulent magnetic-field damping would affect the synchrotron emissivity and thus create narrow structures close to the shock. Alternatively, a very strong magnetic field at the shock would impose strong synchrotron losses and thus limit the width of the rims. 

\subsection{\emph{Model~II:} High damped magnetic field}
\label{subsec:med_mag_fl}

As Model~I involving a weak advected magnetic field fails to reproduce the observed intensity profiles of the synchrotron emission in both the radio
and the X-ray band, we introduce now damping of the turbulent magnetic field. Therefore we set magnetic-field profile described by Eq.~\ref{Bdamp}.
Obviously, in this scenario the magnetic field strength in the
immediate downstream of the forward shock must be larger than for Model~I, because damping suppresses synchrotron emission from the far-downstream region. Since inverse Compton radiation is produced wherever high-energy electrons reside, an overproduction of photons in the $\gamma$-ray band would arise unless the magnetic field is scaled up. Especially when combined with stochastic re-acceleration as an explanation for the radio data, the magnetic-field damping requires an extensive increase of the overall magnetic field. As reflected by the decreasing
width of the turbulent region, the stochastic re-acceleration of particles is more efficient at the early stages of SNR. As the re-accelerated electrons are advected away from the shock,
they mostly experience the damped magnetic field. While the total synchrotron emission depends on the absolute values of the immediate post-shock field strength, $B_2$, and the residual level $B_0$, the shape of the radio filaments is determined (apart from the damping scale $l_\mathrm{d}$) by the ratio of $B_2$ and $B_0$. We find that the immediate post-shock field strength and the residual level must be at least $B_2=330\,\mathrm{\mu}$G (amplification factor $\alpha=20$) and $B_0=70\,\mathrm{\mu G}$, in order to maximize the filling factor, and thus to produce sufficient overall synchrotron emission, and to fit the radio filaments simultaneously. Model~II is based on these minimum values, and results are displayed as green dashed lines in  Figs.~\ref{fig:radio_planck1} and
\ref{fig:xray_emiss}, as well as in the bottom panels of Figs.~\ref{fig:gamma1} and \ref{fig:xrprof1}.

Particle acceleration and propagation is modeled following the same procedure as in Model~I, but some
of the parameters are slightly adjusted.
The second row of Table~\ref{Theor_param} lists the relevant model parameters, which now include the damping length, $l_\mathrm{d}$, and the residual field level, $B_0$. Note, that the damping length, $l_\mathrm{d}$, provides spatial characteristics for the turbulence responsible for the magnetic-field amplification. Likewise, the width of the re-acceleration region, $L_\mathrm{fm}$, represents the damping scale for the fast-mode waves. As both quantities are associated with different types of turbulence, they are independent of each other. At 440~years the amplification efficiency for the second model is $\sim14$.

Along the shock surface certain variations in the parameter values are to be expected, as not all filaments are the same \citep[cf.][]{2014ApJ...790...85R}. Also, Tycho is not a perfectly spherically-symmetric SNR and the projection effects may impose a bias. Therefore our choice for the residual field level and the damping scale rather provide a reasonable
order of their magnitudes as in this work we consider only one particular rim of Tycho. Fitting the \textit{Suzaku} data with an enhanced post-shock magnetic-field $B_2=330\,\mathrm{\mu G}$ requires a moderate increase of the gyrofactor compared to Model~I ($\xi=16$).

Damping quenches the synchrotron emissivity in the deep downstream region. While the suppression is achromatic for power-law electron spectra, i.e. at frequencies below the cutoff in the synchrotron spectrum, it becomes progressively stronger beyond the roll-off energy. A competing process for quenching the emissivity at the roll-off frequency is energy losses preventing the propagation of electron from the shock to the deep downstream region. The synchrotron loss time can be expressed in terms of the magnetic-field strength and the energy of photons, $E_\mathrm{sy}$, that the electrons would typically emit
\begin{equation}
\tau_\mathrm{loss}\simeq 50\,\mathrm{yr}\,\left(\frac{B_2}{100\,\mu\mathrm{G}}\right)^{-3/2}
\,\left(\frac{E_\mathrm{sy}}{\mathrm{keV}}\right)^{-1/2}\ .
\end{equation}
The distance electrons travel during the loss time roughly equals the thickness of the filaments. The particle transport is governed by diffusion and advection. The latter process dominates at lower energies and its length scale is given by  
\begin{align}
 l_\mathrm{adv}&= u_2 \tau_\mathrm{loss}=\frac{V_\mathrm{sh}}{r_\mathrm{sh}}\tau_\mathrm{loss}\nonumber\\
 &\simeq 2 \times10^{17}\,\mathrm{cm}\,\left(\frac{V_\mathrm{sh}}{5000\,\mathrm{km/s}}\right)\left(\frac{B_2}{100\,\mu\mathrm{G}}\right)^{-3/2} \left(\frac{E_\mathrm{sy}}{\mathrm{keV}}\right)^{-1/2}\,.
\label{eq:ad_len}
\end{align}
At high energies diffusion with Bohmian energy scaling becomes the dominant propagation process.
The corresponding distance is energy-independent and given by
\begin{align}
 l_\mathrm{diff}&=\sqrt{D_r\tau_\mathrm{loss}}\nonumber\\
 & \simeq 10^{17}\,\mathrm{cm} \sqrt{\xi}\,\left(\frac{B_2}{100\,\mu\mathrm{G}}\right)^{-3/2}\,.
\label{eq:diff_len}
\end{align}
Equating Eqs.~\ref{eq:ad_len} and \ref{eq:diff_len} one can find the critical photon energy, where the transition from the advection-dominated into diffusion-dominated transport occurs 
\begin{equation}
E_\mathrm{sy,c}\simeq \frac{4\,\mathrm{keV}}{\xi}\left(\frac{V_\mathrm{sh}}{5000\,\mathrm{km/s}}\right)^2\,.
\end{equation}
With $\xi=16$ and $V_\mathrm{sh}=4100\,\mathrm{km/s}$ we obtain for Model~II $E_\mathrm{sy,c}\approx0.2$~keV, implying that both, advection and diffusion impact the distance covered by electrons that account for the 1~keV-rims. Estimating the propagation length for the electrons that radiate 1~keV-photons, we find $l_\mathrm{adv}\simeq  3\times 10^{16}\,\mathrm{cm}$ and $l_\mathrm{diff}\simeq 7\times 10^{16}\,\mathrm{cm}$, and hence diffusive transport is more important, but advective transport is not negligible. Accounting for both advective and diffusive terms, the total transport length that the electrons travel before expending their energy is then given by~\citep{2006A&A...453..387P}
\begin{equation}
 l_\mathrm{loss}=\left(\sqrt{\frac{u_2^2}{4D_r^2}+\frac{1}{D_r\tau_\mathrm{loss}}}-\frac{u_2}{2D_r}\right)^{-1}\simeq 10^{17}\, \mathrm{cm}\,,
\label{eq:tot_len}
\end{equation}
where the last expression applies to electron emitting 1-keV photons and Model~II. The underlying assumption of a constant magnetic-field strength is violated though, as the synchrotron energy losses decrease over a length scale $5\cdot 10^{16}\, \mathrm{cm}$. Therefore the effective loss-length of electrons is much larger than $10^{17}\, \mathrm{cm}$, and the synchrotron filaments are largely shaped by magnetic-field damping.

To account for the radio profiles magnetic-field damping is clearly needed. Radio-emitting electrons have energy-loss times far in excess of the age of SNRs, and the radio rims cannot arise from synchrotron losses. The model with magnetic-field damping (Model~II) fits the spectral data reasonably well. To be noted from Fig.~\ref{fig:xrprof1} is that an effective damping length of
$l_\mathrm{d} = 0.01 R_\mathrm{sh} \simeq 10^{17}\,$cm can indeed reproduce the sharply peaked radio profiles in the shock vicinity, but somewhat underpredicts the radio intensity in the deep downstream region where Rayleigh-Taylors fingers from the contact discontinuity may provide magnetic field amplification \citep{JunNorm95, Bjorn2017} that is not included in our model. In contrast, a radio emission profile calculated for non-damped magnetic field (grey dotted line in bottom panel of Fig.~\ref{fig:xrprof1}) obviously contradicts the observed data. As a relatively high magnetic-field value is required to maintain the total radio and $\gamma$-ray data,
the width of the X-ray rims is inevitably affected by synchrotron losses. Still, magnetic-field damping is more important for the formation of the X-ray filaments since $l_\mathrm{loss}> l_\mathrm{d}$. Without magnetic-field damping, the synchrotron losses for the post-shock magnetic field $B_2=330\,\mathrm{\mu G}$ are able to produce the thin X-ray rims, but fail to form the radio profiles. When introduced as a natural explanation for the radio profiles, magnetic-field damping moreover becomes the dominant process for the production of the X-ray filaments.  
 
We conclude that magnetic-field damping is essential for the radio filaments. Furthermore, for the minimum downstream magnetic field that can explain the complete observed data, $B_2=330\,\mathrm{\mu G}$,  magnetic-field damping is more important for the X-ray profiles than radiative losses. Nevertheless, for 1-keV emission synchrotron losses provides a subordinate additional process for X-ray filaments formation. Simultaneously the width of the radio rim is solely determined by the magnetic-field damping.
In agreement with our results, the joint X-ray and radio analysis of \citet{2015ApJ...812..101T} finds magnetic-field damping as the preferable scenario. Both studies find similar damping lengths: 1\%-2\% of the SNR radius.

\subsection{Cosmic-ray pressure}
\label{sec:crp}
Next, we verify that the test-particle approximation is valid for Model~II. Therefore we calculate the cosmic-ray pressure, given by
\begin{equation}
 P_\mathrm{cr}(r,t)=\frac{c}{3}\int N(r,p,t)\frac{p^2 dp}{\sqrt{p^2+(m_pc)^2}}\,,
\end{equation}
where $m_p$ is the rest mass of proton and $N(r,p,t)$ the differential proton number density. As mentioned before, the pressure exerted by the non-thermal electrons is negligible in our model.
The relative velocity change of the plasma in the shock rest-frame in a particular region between $r_1$ and $r_2$ caused by the particle pressure is
\begin{align}
\frac{\delta u}{u}=&-\frac{1}{u^2 \rho}\int_{r_1}^{r_2}dx\frac{\partial P_\mathrm{cr}(x)}{\partial x} \nonumber \\[9pt]
&= \frac{P_\mathrm{cr}(r_2)-P_\mathrm{cr}(r_1)}{P_\mathrm{flow}}\equiv\frac{\delta P_\mathrm{cr}}{P_\mathrm{flow}}\,.
\label{eq:delu}
\end{align}
Here $\rho$ denotes the density, $u$ the velocity and $P_\mathrm{flow}=\rho u^2$ the dynamical pressure of the plasma. Eq.~\ref{eq:delu} is universally valid and holds in any region of the plasma flow. 

According to \cite{2010ApJ...721..886K}, the test-particle
approximation is justified as long as the cosmic-ray pressure in the precursor does not exceed 10\% of the shock ram pressure. Consequently, according to Eq.~\ref{eq:delu}
the test-particle regime requires $\delta u/u=\delta P_\mathrm{cr}/P_\mathrm{flow}\le0.1$ in the upstream region.
\begin{figure}[tb!]
\includegraphics[width=0.49\textwidth]{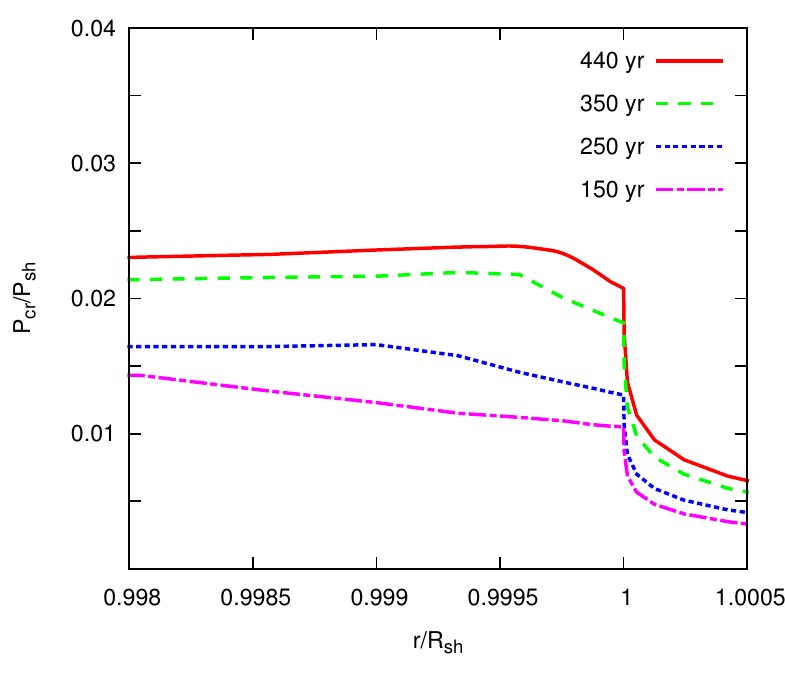}
\caption{Cosmic-ray pressure, $P_\mathrm{cr}$,  normalized by the shock ram pressure, $P_\mathrm{sh}$, for Model~II (damped magnetic field with $B_2=330 \, \mu$G) for different times.}
\label{fig:crp}
\end{figure}
Fig.~\ref{fig:crp} shows the cosmic-ray pressure normalized by the shock ram pressure, $P_\mathrm{sh}=\rho_1 u_1^2$, as a function of the position, $r/R_\mathrm{sh}$, for different times.
To be noted from the figure is that the relative velocity change in the upstream region at 440 years is $\delta u/u=\delta P_\mathrm{cr}/P_\mathrm{sh}\approx0.021$, and hence clearly below the test-particle threshold. Similarly at the earlier stages, the cosmic-ray pressure at the shock does not exceed $\delta u/u<0.1$, verifying that the dynamic cosmic-ray feedback is indeed negligible in our approach.

Behind the shock wave the pressure of particles is significantly enhanced due to stochastic re-acceleration.
In contrast to standard DSA, where the maximal cosmic-ray pressure occurs at the shock, in our model it reaches its maximum in the immediate downstream region.
The corresponding position is determined by the current width of the turbulent zone where the fast-mode waves operate.
Taking into account that the dynamic pressure in the post-shock region is a factor 4 lower than the shock ram pressure in the shock rest-frame, we estimate for the total velocity change at 440~years: $\delta u/u\approx0.01$. We neglect the dynamical feedback from the re-accelerated particles in this work. Nevertheless, this effect may be of interest for future investigations. Indeed, when strong enough, the stochastic acceleration can have significant effect on the post-shock structure. The cosmic-ray pressure in the immediate downstream region enhanced by the stochastic acceleration has to press the plasma away from the re-acceleration region and force it to accumulate directly behind the shock. This process increases the plasma density and the associated magnetic-field strength at the shock of the SNR. It is
clear that this phenomena fundamentally differs from the standard dynamical reaction of particles described in the NLDSA theory.

Summarizing, we conclude that we find a viable self-consistent scenario for Tycho, represented by Model~II.

\subsection{Hadronic model}
\label{modc}

In the previous subsection we presented a lepto-hadronic scenario that requires the lowest possible magnetic field in the remnant compatible with radio filaments and the SED.
More efficient magnetic-field amplification suppresses the inverse-Compton component until the $\gamma$-ray spectrum becomes purely hadronic. In our approach, where we use magnetic-field damping to fit the radio filaments, the magnetic-field values have to be at least $B_0\approx120\,\mu$G and $B_2\approx600\,\mu$G. The total magnetic energy in the remnant is still moderate, $\sim 1.1\times 10^{49}$~erg, on account of the magnetic-field damping.

As discussed in Sec.~\ref{sec:mag_fl}, the separate treatment of magnetic field and hydrodynamics for the ambient density of $n_\mathrm{H}=0.6\,\mathrm{cm}^{-3}$ is justified for post-shock magnetic field below 400~$\mathrm{\mu G}$. Thus, for the hadron-dominated case the dynamical feedback from the magnetic field becomes significant. The corresponding pressure, with the Alfv\'{e}nic Mach number $M_\mathrm{A}\approx9.6$,
would lower the shock compression ratio to $r_\mathrm{sh}\approx 3.87$ and thus soften the particle and radio spectra indices to $s \approx 2.05$ and $\sigma \approx 0.52$, respectively. As the electron spectrum becomes slightly softer on account of magnetic-field impact, the contribution from the stochastic re-acceleration becomes less necessary. Hence, in general, as competing explanation for the softening of the spectra, very high magnetic field inevitably decreases the energy fraction converted to turbulence, $\epsilon$.

Due to the very high magnetic field, X-ray filaments at 1~keV would be primarily governed by the radiative losses, with corresponding propagation scale  $l_\mathrm{loss}\simeq 4\times10^{16}\, \mathrm{cm}$. Thus, the brightness profile would lack the X-ray flux due to extreme synchrotron losses. Nevertheless, as pointed out above, the radio as well the X-ray rims can strongly vary with position along the perimeter due to the natural asymmetry of the remnant. As the electrons emitting the synchrotron radiation in the radio range do not experience
significant losses, the radio profiles do require magnetic-field damping.

Summarizing, we conclude that a purely hadronic model is possible for Tycho, but requires an elaborate and cautious treatment that among other effects includes the dynamical feedback from magnetic-field pressure. Nevertheless, the lepto-hadronic scenario, referred to as Model~II, is able to explain the broadband observations of Tycho in a satisfactory manner.

\section{Summary and conclusion}

In this work we have conducted extensive multiwavelength modeling of Tycho. For the very first time we accounted for stochastic re-acceleration in the downstream region of the forward shock, which provides a consistent explanation to the soft particle spectra without resorting to Alfv\'enic drift \citep{2008A&A...483..529V, 2012A&A...538A..81M, 2014ApJ...783...33S} and its inherent problems, or deducing the particle spectral index from the radio observations \citep{2012ApJ...749L..26A, 2013MNRAS.429L..25Z, 2014NuPhS.256...89C}. As discussed in the introduction, we find the concept of Alfv\'enic drift contradictory, although it has been widely used in the global models for various SNRs. We presented instead a new approach that adds diffusion in momentum space to the standard DSA approach. Furthermore, in our method \textit{both hydrodynamics and particle acceleration are fully time-dependent}. 

The stochastic acceleration of particles in the immediate downstream region is assumed to arise from the fast-mode turbulence, which is supplied by the energy of the background plasma. We found that 2.7\% of the thermal energy density of the downstream background plasma is sufficient to explain Tycho's
soft radio spectra. Simultaneously, the magnetic field is assumed to be amplified by streaming instabilities \citep{BellsInst, Luc&Bell} or turbulent dynamos \citep{2007ApJ...663L..41G}. Based on this, we have presented a self-consistent global model (Model~II) that is able to accurately reproduce the observed radio, X-ray, and $\gamma$-ray emission, and simultaneously account for the non-thermal filaments in radio and X-ray range. The radio filaments are generated due to magnetic-field damping, which is widely considered to allow a relatively low magnetic-field value inside a remnant.
Combining this scenario with stochastic re-acceleration, we found that the minimum magnetic-field required to explain the entire observed dataset of Tycho is $B_2 \approx 330 \, \mu \mathrm{G}$. This value is similar to the results of \cite{2012A&A...538A..81M}, \cite{2008A&A...483..529V} and \cite{2013MNRAS.429L..25Z} although the underlying physical assumptions are quite different. We find that for this minimum magnetic-field strength of $B_2 \approx 330 \, \mu \mathrm{G}$, the X-ray filaments at 1~keV are primarily produced by magnetic-field damping, while the synchrotron losses play a secondary role. This finding is inconsistent with the work of \cite{2012A&A...538A..81M}, who concluded that X-ray filaments shaped by magnetic-field damping are not possible for Tycho. An important criterion here is that {\em the propagation length and hence synchrotron loss scale for electrons radiating at 1~keV is dominated by diffusion}. Therefore, the diffusive transport of particles, which has been previously neglected in all global models of Tycho, has to be taken into account to adequately model the X-ray filaments. We stress that an accurate modeling of the filaments is seamlessly tied to determination of the post-shock magnetic field. Magnetic-field damping is additionally needed for its unique capability to explain the radio rims. In our model for Tycho the magnetic-decay length is of the order of $l_\mathrm{d}\sim 0.01 R_\mathrm{sh} $, which is consistent with the estimate of \cite{2015ApJ...812..101T}. The total magnetic energy in the remnant for Model~II is $3.4\times 10^{48}$~erg. This value is moderate on account of magnetic-field damping, in spite of the efficient magnetic-field amplification at the shock. 

In the framework of our model we predict relatively inefficient Bohm diffusion, reflected in the value of the gyrofactor $\xi\approx 16$. This value is required to consistently fit the synchrotron cutoff observed in the X-ray range. In line with most previous works on Tycho, we also conclude that acceleration of protons is required to explain the $\gamma$-ray flux observed by VERITAS and \textit{Fermi}-LAT. Our research cannot account for a purely leptonic origin for the $\gamma$-ray emission as in \cite{2012ApJ...749L..26A}.
We favor instead a mixed model with a composite flat spectrum ($\Gamma \approx 2$). The electron-to-proton ratio for Model~II is $K_\mathrm{e/p}\approx1/80$ and the maximum energy for protons is $E_\mathrm{max,p}\approx 10 \, \mathrm{TeV}$, since it is linked to the roll-off energy of the synchrotron emission via gyrofactor, $\xi$. This result falls below previously presented $\sim$500~TeV~\citep{2012A&A...538A..81M} and $\sim$50~TeV~\citep{2014ApJ...783...33S}. The maximum energy of electrons is $\sim$3~TeV, which is similar to the value 5 - 6~TeV suggested by \cite{2013MNRAS.429L..25Z} and \cite{2014NuPhS.256...89C}. The total energy in protons in our model is $E_\mathrm{tot,p}\approx2.7\times10^{49}\,\mathrm{erg}$, implying that a few percent of the explosion energy went into cosmic rays. This value is significantly below the 16\% claimed by~\cite{2014ApJ...783...33S}. As a relatively marginal energy fraction is transferred to particles and the cosmic-ray pressure at the shock does not exceed 2.1\% of the shock ram pressure during the entire evolution of the remnant, we can use the test-particle approximation. 

Furthermore, we explicitly show that NLDSA effects are not required, neither to explain the hydrodynamic structure of Tycho nor to produce its slightly concave radio spectrum. First, the hydrodynamic simulations with an ambient gas density of $n_\mathrm{H}=0.6\,\mathrm{cm}^{-3}$ and canonical explosion energy of $10^{51}$~erg provide a decent fit to the observed radii and a reasonable remnant distance $\sim 2.9$~kpc. Second, the radio spectrum is produced by  synchrotron emission generated by electrons that are re-accelerated in the immediate downstream region. In general, the imprint left by stochastic re-acceleration is more prominent in electrons than in protons. Thus, future $\gamma$-ray observations that can successfully discriminate between leptonic and hadronic models for various SNRs may also be able to distinguish between NLDSA and stochastic re-acceleration scenarios.   

We find that a purely hadronic model may also be possible for Tycho for the immediate post-shock magnetic-field $B_2 \approx 600 \, \mu \mathrm{G}$ and far-downstream field $B_0 \approx 120 \, \mu \mathrm{G}$. However, we do not explicitly model the hadronic scenario in this work, because the dynamical reaction of the magnetic field has to be taken into account, an effect that is not yet included in our method. Nevertheless, we favor a lepto-hadronic scenario (Model~II) as it might better produce the X-ray filaments. Additionally, extremely efficient magnetic-field amplification is not required, as in the case of a purely hadronic model.

In our model for Tycho, the contribution of the stochastic re-acceleration in the proton spectrum is marginal due to the small energy fraction converted into downstream turbulence. For a larger amount of energy in fast-mode waves, the corresponding $\gamma$-ray spectrum has to be naturally softer than predicted by the standard DSA, which can provide a viable explanation for the observed soft $\gamma$-ray spectra of some SNRs. This subject is beyond the scope of this work, but could be of interest for future investigations.

Finally, we emphasize that the dynamical feedback on the background plasma from the cosmic rays re-accelerated in the immediate downstream region of the SNR is of great interest for future studies, since it clearly differs from the
classical non-linear effects discussed in the literature.

Summarizing, we find that stochastic re-acceleration is indeed a promising and natural alternative to explain the soft spectra of Tycho, and potentially other SNRs.

\begin{acknowledgements}
We are very grateful to Asami Hayato for kindly providing us the \textit{Suzaku} data.
Furthermore, we acknowledge NASA and \textit{Suzaku} grants, and support by the Helmholtz Alliance
for Astroparticle Physics HAP funded by the Initiative and Networking Fund of the Helmholtz Association. VVD is supported by NSF grant 1911061.
\end{acknowledgements}


\bibliographystyle{aa}
\bibliography{References}

\end{document}